\documentclass{article}
\usepackage[utf8]{inputenc}
\usepackage[top=3cm, bottom=3cm, left=3cm,right=3cm]{geometry}
\usepackage{hyperref}
\usepackage[backend=biber,style=authoryear,natbib=true,
maxcitenames=2,mincitenames=1,uniquelist=false]{biblatex}
\usepackage{graphicx}
\usepackage{color}
\usepackage{amsmath}
\usepackage{amsfonts} 
\usepackage{setspace}
\usepackage{amsmath}
\usepackage{algorithm}
\usepackage{algorithmic}
\usepackage{mathtools}
\usepackage{bbm}
\usepackage{float}
\usepackage{graphicx}
\usepackage{caption}
\usepackage{subcaption}
\usepackage[super]{nth}

\newcommand{\params}{\boldsymbol{\omega}}

\makeatletter
\newcommand{\customlabel}[2]{
\protected@write \@auxout {}{\string \newlabel {#1}{{#2}{}}}}
\makeatother
\usepackage{pifont}

\begin{document}
{\Large Detection of local recombination and mutation hotspots in bacterial genomes using neural posterior estimation}

\vspace*{2cm}
Haochu Liu$^{1,2}$, Richard G. Everitt$^{1,2}$, Xavier Didelot$^{1,2,3,*}$

\vspace*{2cm}
$^1$ Department of Statistics, University of Warwick, United Kingdom\\
$^2$ The Zeeman Institute for Systems Biology and Infectious Disease Epidemiology Research, University of Warwick, United Kingdom\\
$^3$ School of Life Sciences, University of Warwick, United Kingdom\\
$^*$ Corresponding author. Tel: 0044 (0)2476 572827. Email: \verb+xavier.didelot@warwick.ac.uk+

\newpage
\section*{ABSTRACT}
\textbf{Motivation:}
Recombination and mutation rates in bacteria vary along the genome, and regions of elevated rate (hotspots) often reflect selection on specific genes. Detecting them is difficult because the two processes leave overlapping signals in sequence data, clonal relationships act as a confounder, and inference must scale to whole genomes. Existing methods assume a constant mutation rate and/or rely on a small set of summary statistics.

\textbf{Results:}
We present a simulation-based framework that infers gene-specific recombination and mutation rates under a well-specified model of bacterial recombination. Neural posterior estimation (NPE) is trained on simulations conditioned on a fixed clonal genealogy, using a wide range of summary statistics. Gene length is also used as both a parameter and an observed statistic, so that the model learns about its effect on other statistics and thus becomes applicable across different gene lengths. To identify hotspots, we introduce a circular posterior-propagated Difference-of-Gaussians detector that contrasts local and regional smoothers of the posterior log rates, carrying NPE uncertainty into hotspot probabilities. On simulated data, the method recovers the true rates, with credible intervals that narrow as gene length increases. We also apply the new methodology to three real datasets in \textit{Bacillus cereus}, \textit{Staphylococcus aureus} and \textit{Escherichia coli} to showcase its usefulness in a range of situations.

\textbf{Availability and implementation:} The Python implementation is freely available from \url{https://github.com/haochu-liu/bacnpe}.

\newpage
\begin{refsection}

\section{INTRODUCTION}

When bacteria recombine, a relatively small fragment of a donor genome becomes integrated into the genome 
of a recipient, so that the effect of recombination in bacteria is similar to gene conversion in 
eukaryotes rather than crossing-over \citep{Didelot2010c}. 
The ancestral process  for a sample of bacterial genomes is therefore characterised by the ancestral 
recombination graph (ARG) with gene conversion \citep{Wiuf2000}. Inferring the events that took place in 
an ARG model to give rise to a given sample of genomes is a notoriously difficult problem, and not
especially rewarding because genetic data are only weakly informative about specific events, so 
that there would be much uncertainty in the results even if exact inference could be performed
\citep{nielsenInferenceApplicationsAncestral2025}. A simpler problem is to infer the rate of recombination,
which is correlated with a number of well studied summary statistics such as linkage disequilibrium
\citep{Slatkin2008} and homoplasy \citep{maynard1998detecting}. 

The rate of recombination in bacteria varies significantly from one species to another \citep{Vos2009},
but also sometimes from one lineage to another of the same species \citep{Didelot2012b}
or even at a finer resolution depending on specific evolutionary conditions \citep{Didelot2013a}.
It also varies along the genome, with some regions being more likely to recombine than others,
often as a result of selective pressures on specific genes \citep{Yahara2013,Yahara2016}.
Inferring the variations of the recombination rate along the genome is therefore an important goal
in bacterial population genomics, but there are a number of issues to resolve which we investigate 
in this article. Firstly, it is not clear which summary statistics are best to use as the basis of inference. 
Secondly, potential confounding factors such as the clonal relationships between the genomes or potential
variations in the mutation rate along the genome need to be accounted for. Thirdly, inference needs
to be computationally scalable since whole genome sequencing is nowadays standard for bacteriology 
\citep{Didelot2012a,Loman2015}.

Here we present a new inferential framework to infer local recombination and mutation rates and 
identify hotspot candidates from bacterial genomes.
Our methodology is based on a previously proposed approximation to the ARG with gene conversion
known as the ClonalOrigin model \citep{didelot2010inference}. This model has the advantage to be built
around and therefore provide control over the clonal genealogy, which is the tree obtained by 
following the clonal ancestral relationships between the genomes \citep{milkman1990molecular,Didelot2007b}. 
We extend this model straightforwardly by considering that there are several regions along the 
genome, each of which has its own recombination and mutation rate. We call these regions genes
even if they may not represent the biological gene units.
We use a simulation-based approach to inference which exploits the fact that this model is easy
to simulate, but otherwise difficult to perform full inference under \citep{didelot2010inference,Medina2023}.
This approach is related to a previous attempt \citep{ansari2014inference} using approximate 
Bayesian computation (ABC) \citep{tavare1997Inferring, sisson2018handbook}.
However, here we use neural posterior estimation (NPE) \citep{papamakarios2016fast, lueckmann2017flexible, greenberg2019automatic}, which has several
advantages compared to ABC in this context. In particular, it allows us to use a much larger set of
summary statistics, without worrying about their relative informativeness. Furthermore, we train the model
across a whole range of gene lengths, using this as both a parameter and an input of the model.
Consequently, the trained estimator learns about the effect that gene length has on the summary statistics and
their relation with the local recombination and mutation rates. Once trained, the model can be applied
to each gene separately, making it computationally scalable.
To identify hotspots, we introduce a circular posterior-propagated Difference-of-Gaussians detector that contrasts local and regional-background smoothers of the posterior log rates. This procedure accounts for circular genome structure and propagates posterior uncertainty from NPE into the hotspot probabilities.

Below we describe in full detail our methodology for inference and provide an implementation in Python at
\url{https://github.com/haochu-liu/bacnpe} which should be easily applicable to any alignment of bacterial genomes.
We test the inference thoroughly using simulated datasets in which the correct values of parameters
are known so that we can benchmark the accuracy and precision of inference. We also apply our 
method to three previously described real datasets from \textit{Bacillus cereus} \citep{didelot2010inference}, \textit{Staphylococcus aureus} \citep{Everitt2014}, and \textit{Escherichia coli} \citep{didelot2012impact} to showcase
the insights that can be obtained by inferring variations in the evolutionary rates along these genomes.

\section{MATERIALS AND METHODS}

\subsection{The ClonalOrigin model}

ClonalOrigin is an approximation to the ancestral recombination graph (ARG) with gene conversion, conditional on a fixed clonal genealogy $\mathcal{T}$ \citep{didelot2010inference}. Recombination events are assumed to occur independently along the clonal genealogy. Under this approximation, non-clonal lineages may coalesce only with the clonal genealogy: they neither coalesce with one another nor undergo further recombination. These restrictions are motivated by the small amount of ancestral material typically carried by a non-clonal lineage, which makes additional events on such lineages unlikely to affect the realised genealogy. They also reduce the computational cost of simulation and yield a process that is Markovian along the genome \citep{didelot2010inference,ansari2014inference}.

Let $T$ denote the total branch length of $\mathcal{T}$, let $\rho_s/2$ denote the recombination rate per site and unit of branch length, and let recombination tract lengths follow a geometric distribution with mean $\delta$. The number of recombination events whose imported tracts cover the first site is then
\begin{equation}
    R_1 \mid \mathcal{T},\rho_s,\delta
    \sim
    \operatorname{Poisson}\!\left(\frac{\rho_s\delta T}{2}\right).
    \label{eq:R_1}
\end{equation}
For each recombination event $e$, the departure point $b_e$ is sampled along the clonal genealogy in proportion to branch length, and the conditional density of its arrival point $a_e$ is
\begin{align}
    p(b_e \mid \mathcal{T}) &= \frac{1}{T},
    \label{eq:p_b}\\
    p(a_e \mid b_e,\mathcal{T}) &= \exp\!\left[-L(a_e,b_e)\right],
    \label{eq:p_a}
\end{align}
where $L(a_e,b_e)$ denotes the total branch length of $\mathcal{T}$ between the time of $a_e$ and that of $b_e$.

The Markovian representation updates the set of recombination events as the process moves from site $i$ to site $i+1$. Because geometric tract lengths are memoryless, an event affecting site $i$ remains active at the next site with probability $1-\delta^{-1}$. Consequently, if $R_i$ events affect site $i$, the number that survive to site $i+1$ satisfies
\begin{equation}
    R_{i+1}^{\mathrm{S}}
    \sim
    \operatorname{Binomial}\!\left(R_i,1-\delta^{-1}\right).
    \label{eq:R_s}
\end{equation}
The number of newly initiated events is modelled as
\begin{equation}
    R_{i+1}^{\mathrm{N}}
    \mid \mathcal{T},\rho_s
    \sim
    \operatorname{Poisson}\!\left(
        \frac{\rho_sT}{2}
    \right),
    \label{eq:R_n}
\end{equation}
and the total number of events affecting site $i+1$ is
\begin{equation}
    R_{i+1}=R_{i+1}^{\mathrm{S}}+R_{i+1}^{\mathrm{N}}.
\end{equation}
Together with their arrival and departure points, these events determine the local genealogy at each site.

Let $T_i$ denote the total branch length of the local genealogy at site $i$, and let $\theta_s/2$ denote the mutation rate per site and unit of branch length. Conditional on the local genealogy, the number of mutations at site $i$ is
\begin{equation}
    M_i \mid T_i
    \sim
    \operatorname{Poisson}\!\left(\frac{\theta_s T_i}{2}\right).
    \label{eq:M_i}
\end{equation}
The resulting local genealogies and mutation events define the simulated sequence alignment. The sequential simulation procedure used to generate these alignments is described in Supplementary Text~\ref{appendix:pseudo_code}.

For each simulated genomic segment, we calculated a set of summary statistics designed to retain information relevant to recombination and mutation while substantially reducing the dimensionality of the inputs. Formal definitions of the summary statistics are provided in Supplementary Text~\ref{appendix:summary_stats}.

\subsection{Simulation-based inference}
\label{sec:sbi}

Approximate Bayesian computation (ABC) has long provided a likelihood-free approach by comparing observed and simulated summary statistics and retaining parameter values that produce sufficiently similar data \citep{tavare1997Inferring,sisson2018handbook}. Previous work applied ABC to infer recombination parameters under the ClonalOrigin model \citep{ansari2014inference}. Conventional ABC requires the choice of a distance function and acceptance tolerance, and its computational cost increases when inference must be repeated for many observations. Earlier applications have therefore primarily focused on parameters describing an entire dataset or genomic region, whereas estimating separate recombination and mutation parameters for many genes requires a more scalable approach.

Neural simulation-based inference (SBI) replaces the accept--reject step of ABC with a flexible neural density estimator trained on simulated parameter--observation pairs \citep{lueckmann2021benchmarking}. Depending on the method, the network may approximate the likelihood, a likelihood-to-evidence ratio, or the posterior distribution. Once trained over the prior predictive distribution, the resulting estimator can be reused for new observations without additional simulations. This amortisation is central to our application: a single trained model can provide posterior distributions for the local recombination and mutation parameters of every genomic segment, making gene-level inference computationally feasible.

We use neural posterior estimation (NPE) \citep{papamakarios2016fast, lueckmann2017flexible, greenberg2019automatic}, which directly approximates the posterior distribution. Let $\params=(\rho_s,\theta_s,\ell)$ denote the parameters of a genomic segment, namely its local recombination and mutation parameters together with its length $\ell$, and let $z$ denote the corresponding observation, which contains the summary statistics and the length of the segment (Section~\ref{sec:compression_network}). Parameter--observation pairs are generated by sampling $\params_i\sim\pi(\params)$ and then simulating $z_i\sim p(z\mid\params_i)$. A conditional density estimator $q_{\phi}(\params\mid z)$ is trained by minimising the empirical negative log density of the generating parameters,
\begin{equation}
    \mathcal{L}_{\mathrm{NPE}}(\phi)
    =
    -\frac{1}{N}\sum_{i=1}^{N}
    \log q_{\phi}(\params_i\mid z_i).
    \label{eq:npe_main_loss}
\end{equation}
In expectation, this objective equals, up to a constant independent of $\phi$, the expected Kullback--Leibler divergence between the true and the estimated posterior distributions, so that minimising it drives $q_{\phi}(\params\mid z)$ towards the true posterior. Specifically, we use the NPE-C variant of NPE \citep{greenberg2019automatic}, also known as automatic posterior transformation. Because all of our training simulations are drawn directly from the prior in a single round, no proposal correction is required and the NPE-C loss reduces to Equation~\eqref{eq:npe_main_loss}. The full algorithm is provided in Supplementary Text~\ref{appendix:npe}.

We parameterise the posterior estimator using a conditional neural spline flow (NSF) \citep{durkan2019neural}, which transforms a simple base distribution into a flexible, observation-dependent density over $\params$. The NSF and the compression network described below are trained jointly. After training, posterior samples for an observed segment can be drawn directly from $q_{\widehat{\phi}}(\params\mid z_{\mathrm{obs}})$, avoiding a separate Markov chain Monte Carlo (MCMC) run for each gene. Compared with neural likelihood estimation (NLE; \citealp{papamakarios2019sequential}) and neural ratio estimation (NRE; \citealp{hermans2020likelihood}), this direct sampling makes NPE particularly well suited to repeated local inference. NLE and NRE are described in Supplementary Texts~\ref{appendix:nle} and \ref{appendix:nre}, and their empirical comparison with NPE is reported in Supplementary Text~\ref{appendix:benchmark}. Under the configurations examined here, NPE provided the strongest combination of statistical performance and computational efficiency. All methods were implemented using the \texttt{sbi} package \citep{BoeltsDeistler_sbi_2025}.

\subsection{Compression network}
\label{sec:compression_network}

Each gene was represented by a vector
\begin{equation}
    z = (s_1,\ldots,s_d,\ell) \in \mathbb{R}^{d+1},
\end{equation}
where $s_1,\ldots,s_d$ are the summary statistics listed in Supplementary Text~\ref{appendix:summary_stats}, and $\ell$ is the gene length. We used a fully connected multilayer perceptron (MLP) $g_{\psi_{\mathrm{e}}}:\mathbb{R}^{d+1}\to\mathbb{R}^{m}$, with parameters $\psi_{\mathrm{e}}$, to compress this input into $m$ learned nonlinear features, $\widetilde{z}=g_{\psi_{\mathrm{e}}}(z)$.

To ensure that gene length remained explicitly available to the posterior estimator, we appended its original value to the MLP output:
\begin{equation}
    e_{\psi_{\mathrm{e}}}(z)
    =
    \begin{bmatrix}
        \widetilde{z}\\
        \ell
    \end{bmatrix}
    \in \mathbb{R}^{m+1}.
\end{equation}
The first $m$ dimensions are learned nonlinear summaries of the full input, whereas the final dimension is the gene length itself, passed through unchanged.

This design was motivated by the fact that gene length is directly interpretable and may affect the amount of information contained in the gene sequences. This length-preserving design retains a directly interpretable feature for the downstream density estimator, while allowing the MLP to learn nonlinear interactions between gene length and the other inputs. The resulting embedding conditions the NSF,
\begin{equation}
    q_{\phi}\!\left(\params \mid e_{\psi_{\mathrm{e}}}(z)\right),
    \label{eq:npe_embedded}
\end{equation}
and the flow and embedding parameters, $\phi$ and $\psi_{\mathrm{e}}$, are optimised jointly by minimising Equation~\eqref{eq:npe_main_loss} with $q_{\phi}(\params_i\mid z_i)$ replaced by $q_{\phi}(\params_i\mid e_{\psi_{\mathrm{e}}}(z_i))$. We used rectified linear unit (ReLU) activations. The network architecture and NPE training hyperparameters were selected according to final model performance, as described in Supplementary Text~\ref{appendix:tuning}.

\subsection{Diagnostics}

We first evaluated empirical coverage and Euclidean error on simulated test datasets. Empirical coverage was calculated as the proportion of simulations for which the true parameter value fell within the corresponding $95\%$ posterior credible interval; values below or above the nominal level indicate overconfidence or underconfidence, respectively. Euclidean errors between the posterior median and the true parameter value were used to assess point-estimation accuracy, with smaller values indicating more accurate inference.

We also assessed posterior calibration using simulation-based calibration (SBC) \citep{talts2018validating} and tests of accuracy with random points (TARP) \citep{lemos2023sampling}. SBC checks whether the ranks of parameters drawn from the prior are uniformly distributed within their corresponding posterior samples, and therefore primarily assesses marginal calibration. We quantified departures from uniformity using one-sample Kolmogorov--Smirnov tests on the normalised ranks. TARP complements SBC by evaluating calibration in the joint parameter space through distance-based credible regions. Its expected coverage probability curve should follow the diagonal for a calibrated posterior; departures were summarised using the area between the empirical curve and the diagonal, together with a one-sample Kolmogorov--Smirnov test.

Finally, we compared inferred segment lengths with their true or observed values. Because segment length is supplied directly to the model and retained by the length-preserving compression network, agreement provides a simple additional check that this information is preserved by the inference pipeline and offers an indication of overall model fit.

\subsection{Circular posterior-propagated Difference-of-Gaussians hotspot detection}
\label{sec:ppdog}

After obtaining gene-specific posterior distributions from NPE, we identified localised elevations in recombination or mutation rate using a circular posterior-propagated Difference-of-Gaussians (ppDoG) detector. The method treats the posterior log rates along a chromosome as an irregularly sampled spatial signal. A narrow Gaussian smoother represents the local rate, whereas a broader smoother represents the regional background \citep{marr1980theory,lindeberg1994scale}. Their contrast therefore detects local enrichment at genomic scales determined by the two bandwidths.

For each posterior draw, rates were transformed to the log scale and smoothed using circular genomic distances, so that genes near the origin and end of a circular chromosome were treated as neighbours. Contributions from nearby genes were weighted by both spatial proximity and posterior precision. If $N_i^{(m)}$ and $B_i^{(m)}$ denote the narrow- and broad-scale smoothed log rates for gene $i$ in posterior draw $m$, respectively, the ppDoG score is
\begin{equation}
    D_i^{(m)}=N_i^{(m)}-B_i^{(m)}.
    \label{eq:ppdog_main_score}
\end{equation}
Because the score is calculated on the log scale, $\exp(D_i^{(m)})$ is the corresponding local-to-background fold enrichment.

Given a minimum fold enrichment $F_{\min}>1$, we estimated the posterior hotspot probability from $M$ posterior draws as
\begin{equation}
    \widehat{P}_i^{\mathrm{hot}}
    =
    \frac{1}{M}\sum_{m=1}^{M}
    \mathbb{I}\!\left[D_i^{(m)}>\log F_{\min}\right].
    \label{eq:ppdog_main_probability}
\end{equation}
Genes with $\widehat{P}_i^{\mathrm{hot}}\geq0.9$ were designated candidate hotspot-associated genes. Applying the detector to posterior draws, rather than only to posterior medians, propagates uncertainty from NPE into the hotspot calls. The detector identifies relative spatial enrichment rather than simply selecting genes with the largest absolute rates: a high-rate region may have a weak response if its surrounding background is similarly elevated, whereas a moderate rate may yield a strong response when it forms a distinct local peak. Full definitions of the circular distance, precision-weighted Gaussian smoothers, and bandwidth interpretation, together with the bandwidths and thresholds used in each application, are provided in Supplementary Text~\ref{appendix:ppdog}.

\section{RESULTS}

\subsection{Application to simulated datasets}
\label{sec:application_clonalorigin_sim}

We simulated a clonal genealogy with $30$ leaves, denoted by $\mathcal{T}_{30}$. Let $p(z \mid \params, \mathcal{T}_{30})$ denote the ClonalOrigin simulator, where $z=(s_1,\ldots,s_d,\ell)$ contains the summary statistics of a simulated gene sequence together with its length (Section~\ref{sec:compression_network}), and $\params = (\rho_s, \theta_s, \ell)$ denotes the simulation parameters. We assigned the independent priors
$\rho_s \sim \mathcal{U}(0,0.1)$, 
$\theta_s \sim \mathcal{U}(0,0.1)$ and 
$\ell \sim \operatorname{DiscreteUniform}\{100,\ldots,10{,}000\}$,
where $\rho_s/2$ and $\theta_s/2$ are the recombination and mutation rates per site and unit of branch length, and $\ell$ is the gene length in base pairs (bp). In total, we generated $50{,}000$ simulations for model training and $1{,}000$ independently simulated datasets for testing.

\begin{figure}[H]
    \centering
    \includegraphics[width=15cm]{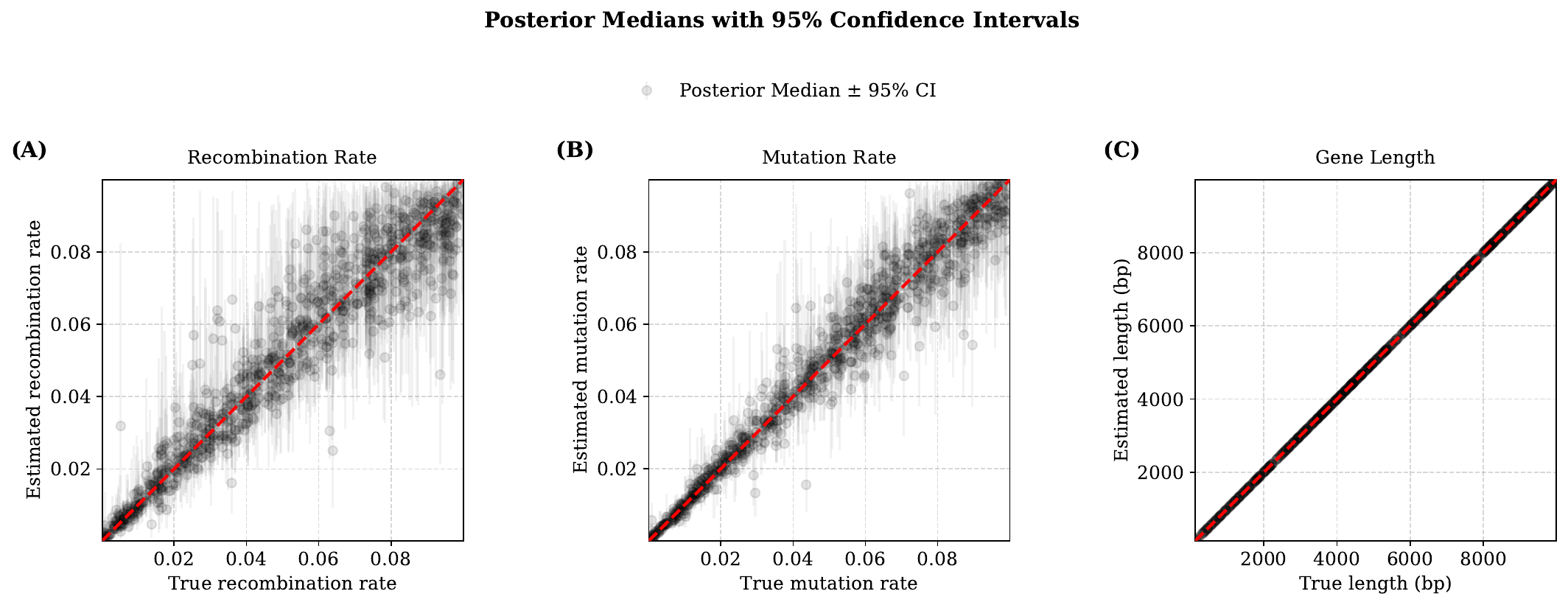}
    \caption{Posterior estimates for the simulated test datasets. Each point represents the posterior median obtained by conditioning the trained NPE model on one test observation, and the vertical interval denotes the corresponding $95\%$ credible interval. The horizontal axis shows the true parameter value used to generate each observation, and the vertical axis shows the inferred posterior estimate.}
    \label{fig:coverage}
\end{figure}

Using the tuned configuration in Supplementary Text~\ref{appendix:tuning}, we trained the NPE-C model on the training set to obtain an amortised approximation to the posterior distribution for each observation. Figure~\ref{fig:coverage} presents the posterior medians and $95\%$ credible intervals for the test datasets, together with the corresponding true parameter values. The gene length is almost perfectly inferred, as would be expected from the fact that it is used as both an input and an output of the method. This provides a first check that the method is behaving as expected, and enables control for the effect of the gene length on other statistics. Both the recombination rate and mutation rate inferred values show good correlation with the correct values, indicating that the data are informative about both parameters and that the method is able to extract that information. Some credible intervals on the mutation rates are wider than others, reflecting comparatively less information, as expected especially for shorter genes. Some credible intervals on the recombination rates are even wider than those for the mutation rates, because in addition to the gene length effect there is an effect of mutation rate, whereby a low mutation rate results in fewer mutations and therefore less information about the recombination rate.

\begin{figure}[htbp]
    \centering
    \begin{subfigure}[b]{0.32\textwidth}
        \centering
        \includegraphics[width=\textwidth]{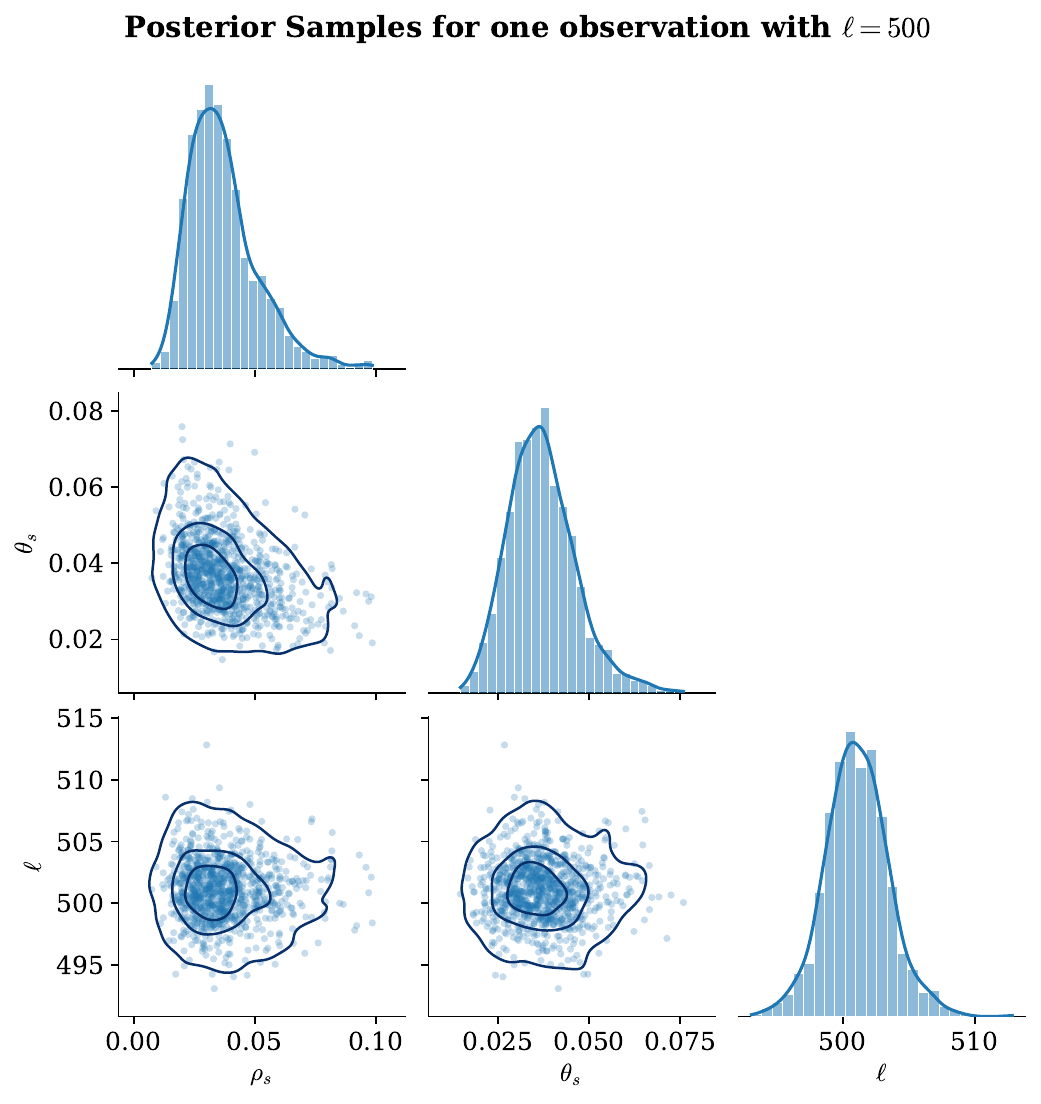}
        \caption{Gene length $500\,\mathrm{bp}$}
        \label{fig:sub1}
    \end{subfigure}
    \hfill
    \begin{subfigure}[b]{0.32\textwidth}
        \centering
        \includegraphics[width=\textwidth]{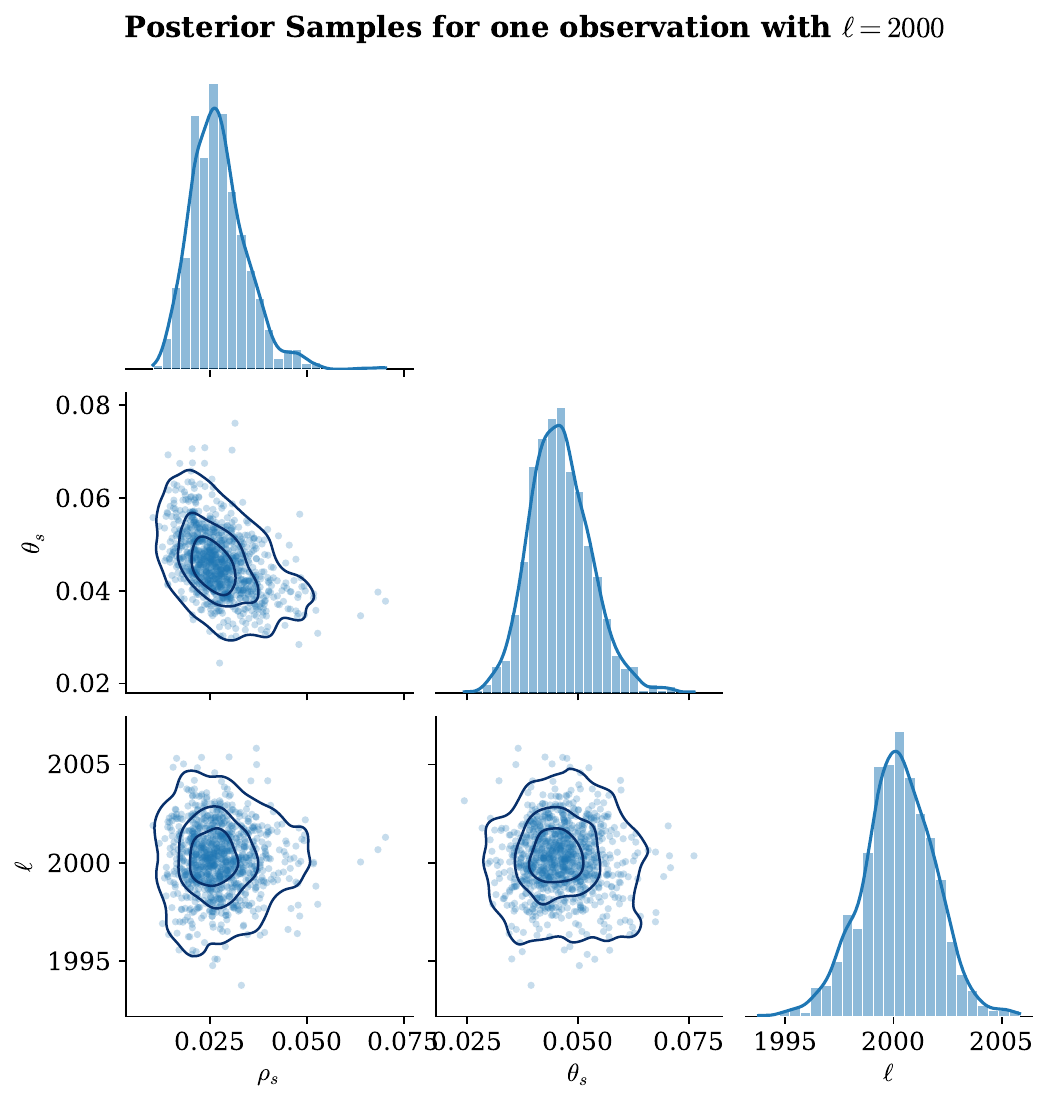}
        \caption{Gene length $2000\,\mathrm{bp}$}
        \label{fig:sub2}
    \end{subfigure}
    \hfill
    \begin{subfigure}[b]{0.32\textwidth}
        \centering
        \includegraphics[width=\textwidth]{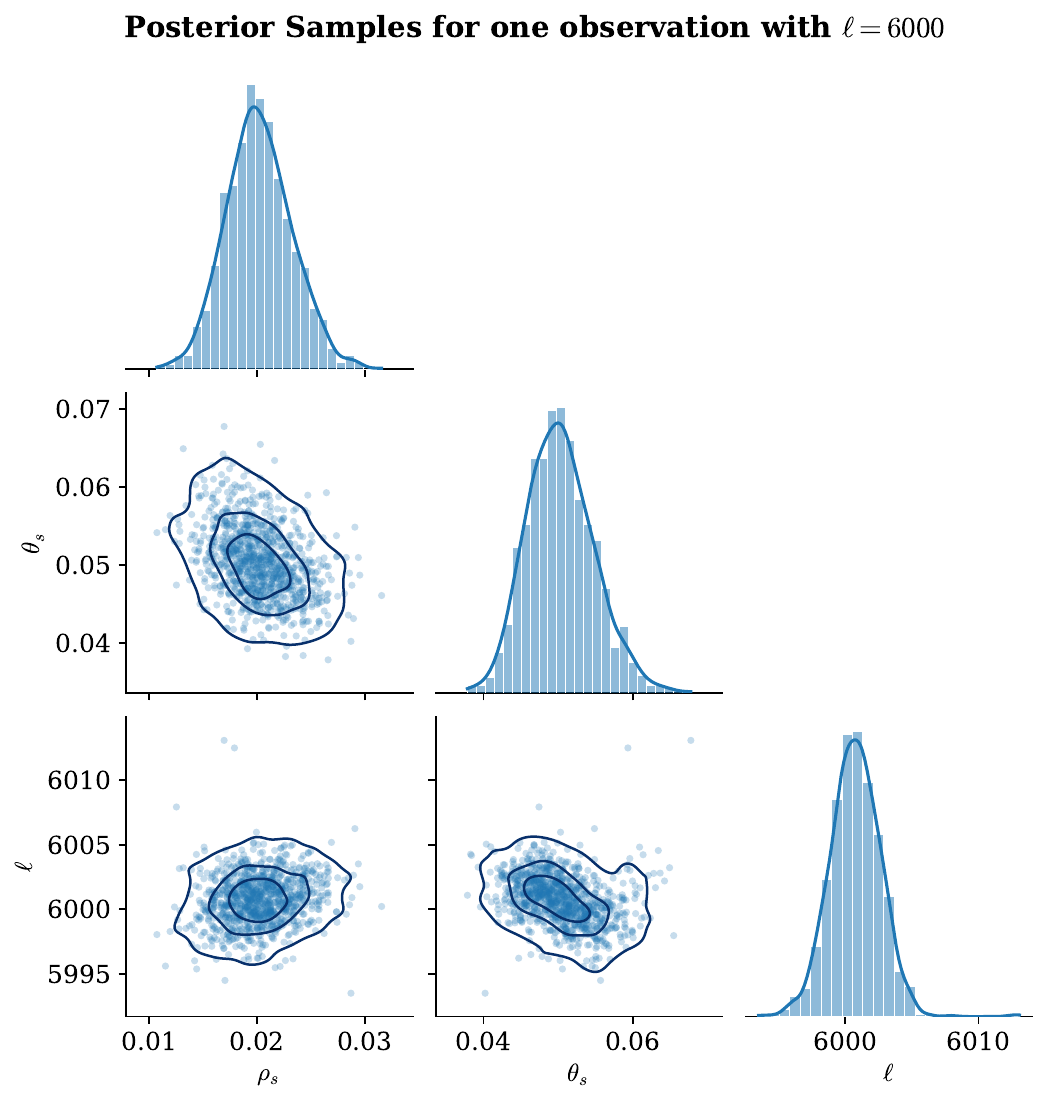}
        \caption{Gene length $6000\,\mathrm{bp}$}
        \label{fig:sub3}
    \end{subfigure}
    
    \caption{Posterior samples for three observations generated by the same recombination and mutation rates, $\rho_s=0.02$ and $\theta_s=0.05$ but different gene lengths}
    \label{fig:sim_diff_L}
\end{figure}

To illustrate the effect of gene length on the inference of the mutation and recombination rates, Figure~\ref{fig:sim_diff_L} compares posterior samples for three representative observations generated using the same recombination and mutation rates, $\rho_s=0.02$ and $\theta_s=0.05$, but different gene lengths of $500$, $2{,}000$ and $6{,}000\,\mathrm{bp}$. The credible intervals for $\rho_s$ and $\theta_s$ become narrower as gene length increases. There is a slight negative correlation between the inferred values of $\rho_s$ and $\theta_s$, especially for the shortest gene (500\,bp), which can be explained by the possibility of recombination coming from above the root of the tree and therefore having a similar effect as mutation on the number of polymorphic sites. However, this effect is small and becomes even smaller for longer genes, demonstrating that in these conditions the set of summary statistics used is able to distinguish between mutation and recombination events.

\begin{figure}[H]
    \centering
    \begin{subfigure}[b]{0.8\textwidth}
        \centering
        \includegraphics[width=\textwidth]{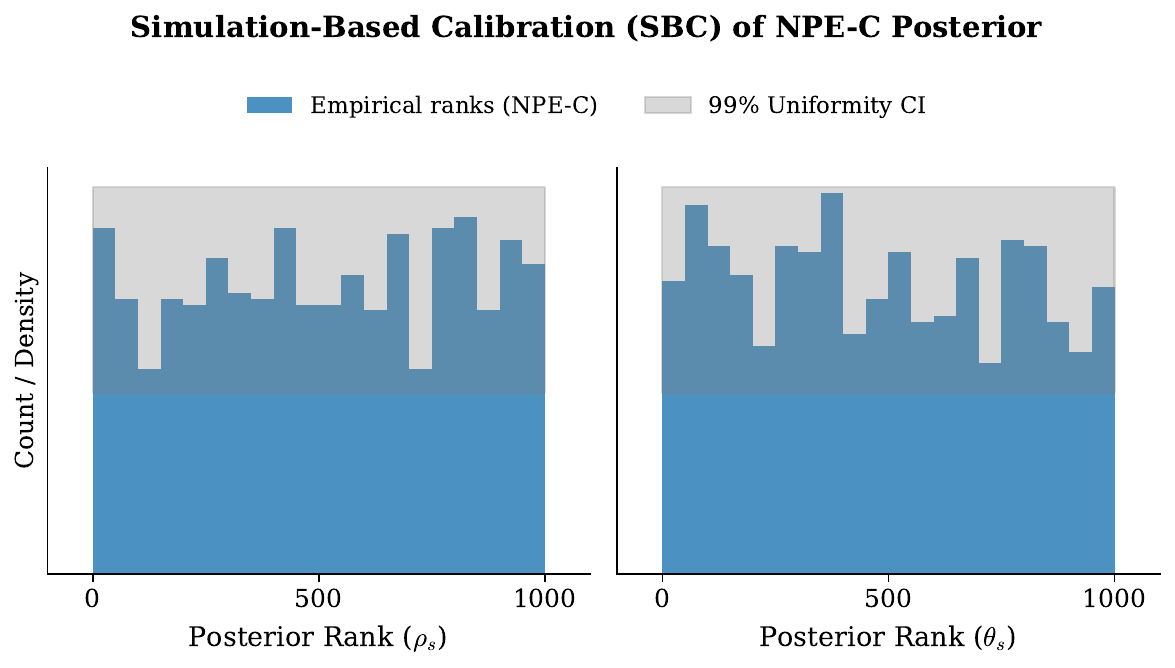}
        \caption{SBC rank histograms for $\rho_s$ (left) and $\theta_s$ (right); the grey bands show the $99\%$ intervals expected under uniformity.}
        \label{fig:sbc_results}
    \end{subfigure}
    \vspace{0.3cm} 

    \begin{subfigure}[b]{0.4\textwidth}
        \centering
        \includegraphics[width=\textwidth]{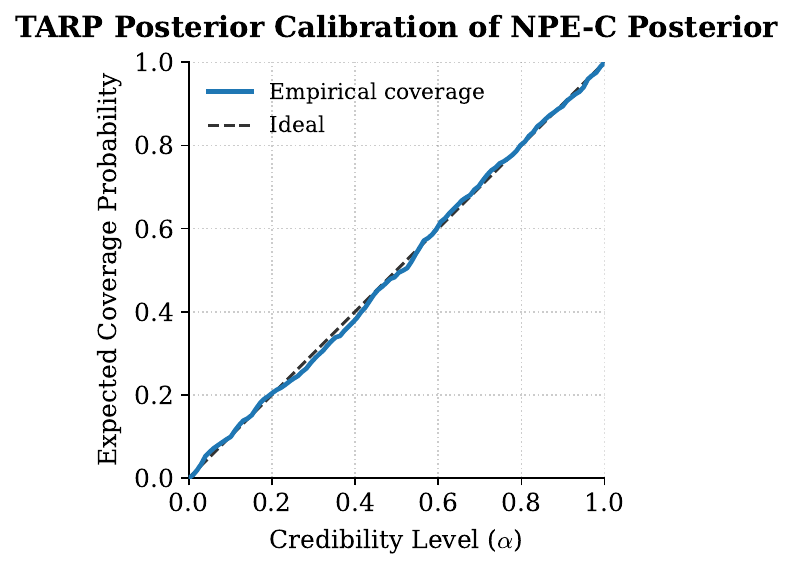}
        \caption{TARP expected coverage probability against the credibility level; the dashed line indicates ideal calibration.}
        \label{fig:tarp_results}
    \end{subfigure}
    \hfill 
    \begin{subfigure}[b]{0.4\textwidth}
        \centering
        \includegraphics[width=\textwidth]{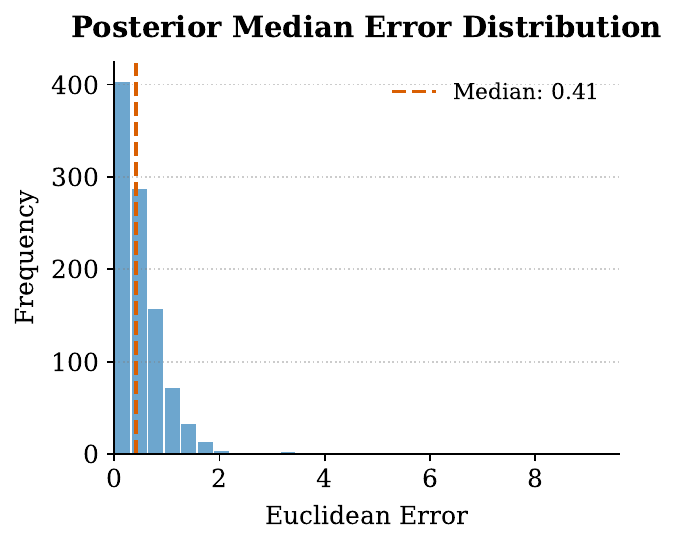}
        \caption{Euclidean errors between posterior medians and true parameter values; the dashed line marks the median error.}
        \label{fig:error_results}
    \end{subfigure}

    \caption{\textbf{Validation and calibration diagnostics for the NPE-C posterior estimator.}
    (\textbf{a}) Simulation-based calibration (SBC) rank histograms for $\rho_s$ and $\theta_s$.
    (\textbf{b}) Tests of accuracy with random points (TARP), comparing the empirical expected coverage probability (solid blue) with ideal calibration (dashed black).
    (\textbf{c}) Distribution of Euclidean errors between posterior medians and the true parameter values; the dashed orange line indicates the median error of $0.41$.}
    \label{fig:diagnostics_results}
\end{figure}

Overall, the diagnostic results indicated that the NPE-C posterior was well calibrated and provided accurate point estimates (Figure~\ref{fig:diagnostics_results}). The empirical coverage of the $95\%$ posterior credible intervals was $93.6\%$ for $\rho_s$ and $95.1\%$ for $\theta_s$, both close to the nominal level. Euclidean errors between the posterior medians and true parameter values were generally small, with a median of $0.41$. The SBC rank histograms showed no pronounced departures from uniformity, with all bin counts lying within the $99\%$ intervals expected under uniformity, while the TARP expected coverage curve closely followed the diagonal, supporting calibration of the joint posterior. Finally, the inferred gene lengths almost perfectly matched their true values (Figure~\ref{fig:coverage}), confirming that gene-length information was preserved by the inference pipeline.

\subsection{Application to the \textit{Bacillus cereus} group}

We reanalysed a dataset consisting of 13 genomes from the 
\textit{Bacillus cereus} group that has been compiled by \citet{didelot2010inference}. 
The \textit{B. cereus} group includes the species \textit{B. cereus}
itself, but also other closely related species, in particular
\textit{B. thuringiensis} which is not phylogenetically distinct from \textit{B. cereus}
as well as \textit{B. anthracis} which is the causative agent of anthrax \citep{Priest2004}.
The genomes were previously aligned into 1218 homologous blocks using 
progressiveMauve \citep{Darling2010} and a genealogy inferred using
ClonalFrame \citep{Didelot2007b}, and we reuse these as the starting point
for our own analysis, as well as the previous estimate of the mean recombination tract length $\delta=236$\,bp \citep{didelot2010inference}. 
To train the dataset-specific inference model, parameter values were sampled independently from the priors
$\rho_s \sim \mathcal{U}(0,0.2)$, 
$\theta_s \sim \mathcal{U}(0,0.1)$ and 
$\ell \sim \operatorname{DiscreteUniform}\{100,\ldots,10{,}000\}$.
For each sampled parameter vector, a gene sequence was simulated conditional on the fixed \textit{B. cereus} genealogy, and the corresponding summary statistics were calculated. The resulting parameter--observation pairs were then used to train the model for this application.
The trained model was conditioned on the observed genomic segments to obtain segment-specific posterior distributions for the local recombination and mutation parameters.
Because NPE provides amortised inference, posterior samples could be generated efficiently for each segment without retraining the model or running a separate MCMC analysis.
As a diagnostic of the inference pipeline, Figure~\ref{fig:bacillus_length} compares the inferred segment lengths with their observed values.

Generating the training data was the most computationally demanding step. Each simulation required on average about 61 s of CPU time. Running 12 simulations in parallel on a single HPC compute node (2 $\times$ Intel Xeon Platinum 8268, 2.9 GHz), 50,000 simulations took approximately 96 h of wall-clock time. Training the NPE model on these simulations required approximately 56 minutes using a single core of an Intel 13th Gen Core i7-1365U CPU (no GPU), with \texttt{sbi} v0.26.1 and \texttt{PyTorch} v2.5.1. Although generating the simulations involved a substantial computational cost, this cost is incurred only once. Because NPE is amortised, the trained model can be applied to all observed segments, as well as to additional observed genes under the same inference setting, without repeating the simulations or retraining the model.

\begin{figure}[H]
    \centering
    \includegraphics[width=15cm]{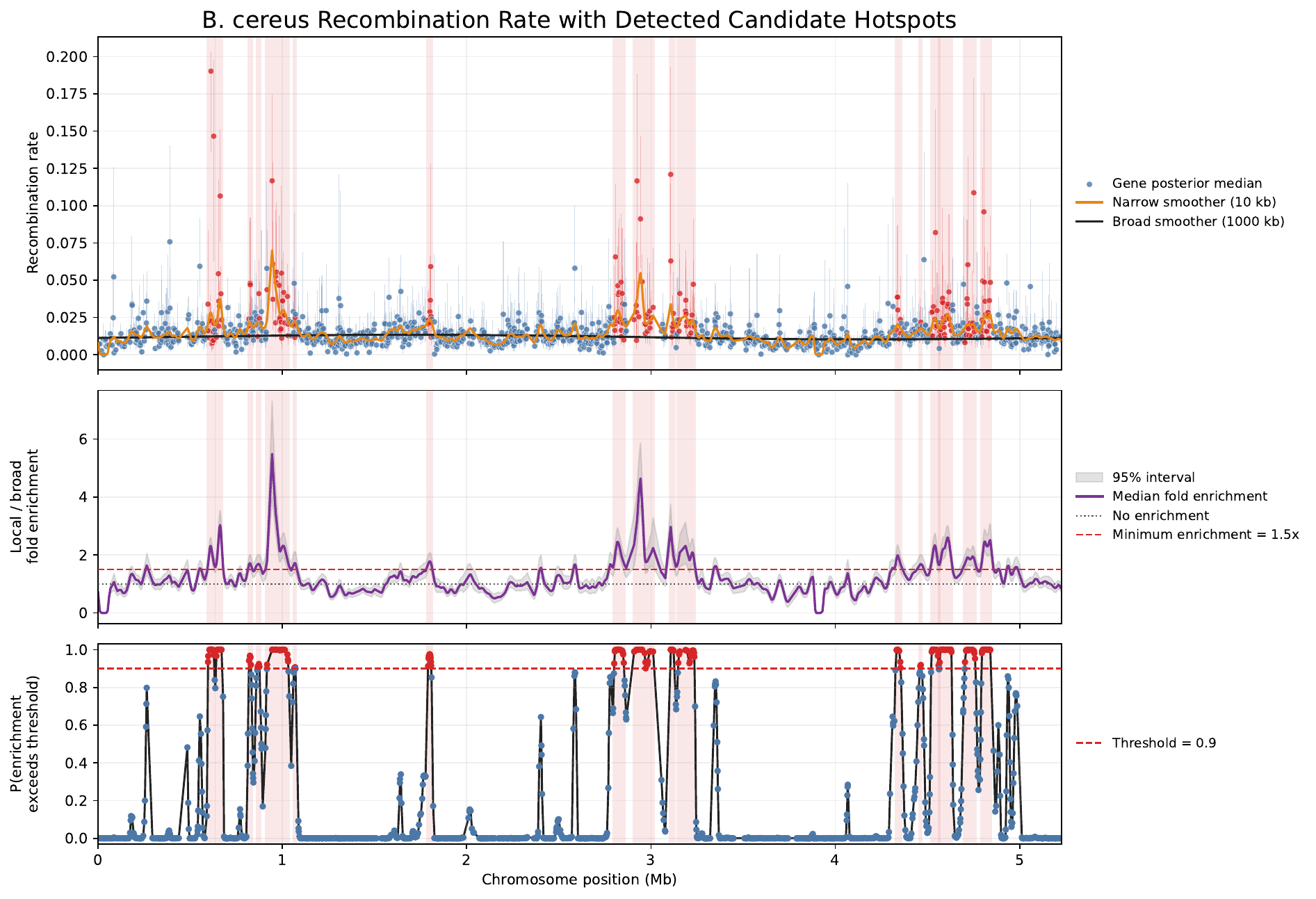}
    \caption{Gene-level posterior estimates of the local recombination parameter across the \textit{B. cereus} genome. Candidate recombination hotspots were identified using ppDoG with $h_{\mathrm{N}}=10{,}000\,\mathrm{bp}$, $h_{\mathrm{B}}=1{,}000{,}000\,\mathrm{bp}$, and $F_{\min}=1.5$.}
    \label{fig:bacillus_recomb}
\end{figure}

\begin{figure}[H]
    \centering
    \includegraphics[width=15cm]{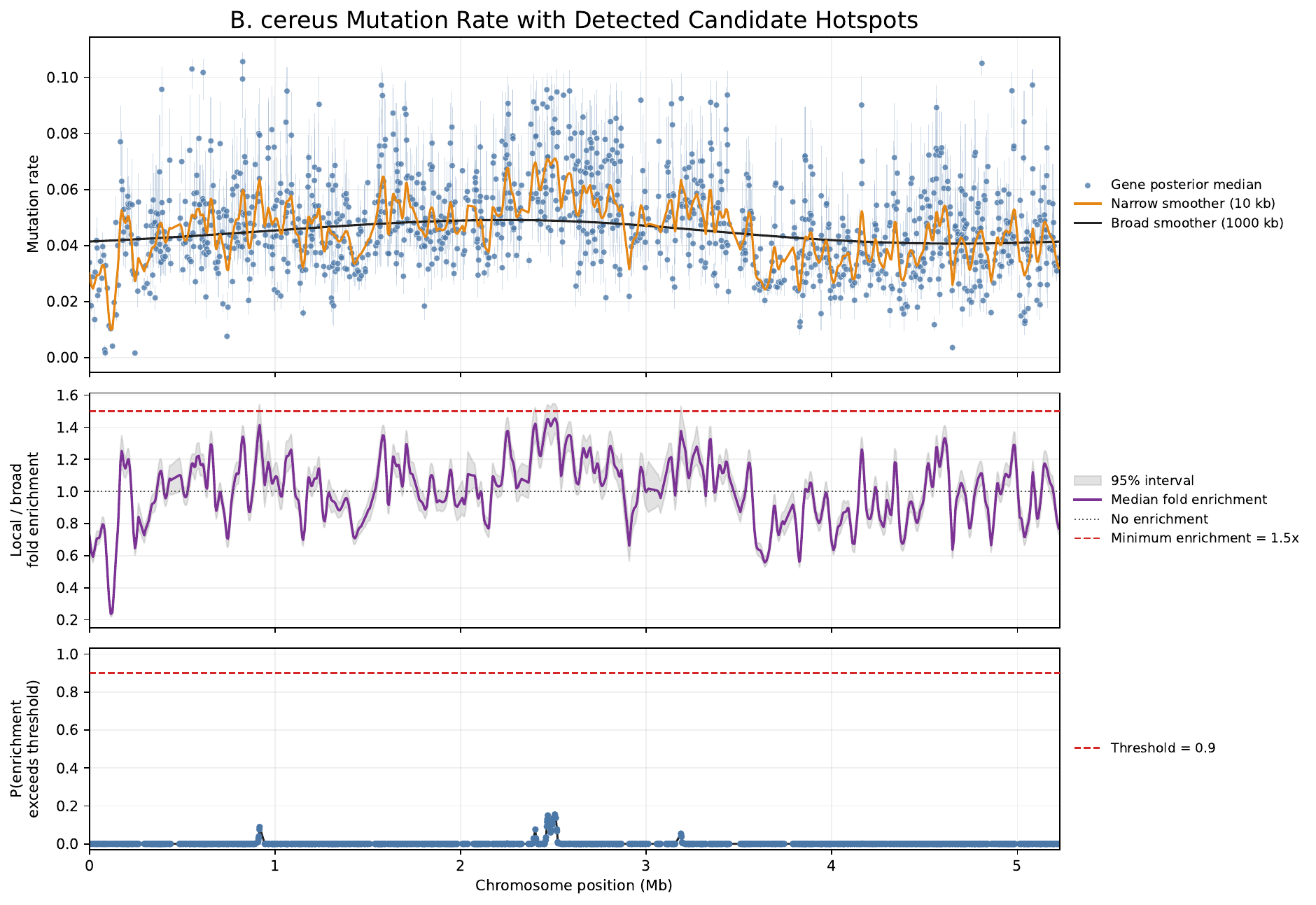}
    \caption{Gene-level posterior estimates of the local mutation parameter across the \textit{B. cereus} genome. Candidate mutation hotspots were identified using ppDoG with $h_{\mathrm{N}}=10{,}000\,\mathrm{bp}$, $h_{\mathrm{B}}=1{,}000{,}000\,\mathrm{bp}$, and $F_{\min}=1.5$.}
    \label{fig:bacillus_mutation}
\end{figure}

Figures~\ref{fig:bacillus_recomb} and \ref{fig:bacillus_mutation} present the recombination and mutation results obtained using ppDoG. 
The coordinates used on the x-axis correspond to positions along
the reference genome \textit{B. anthracis} Ames 0581 for which a 
full annotation has been published \citep{ravelCompleteGenomeSequence2009}.
It is interesting to note that no mutation hotspots were found, but
that in contrast there were 17 recombination hotspots detected, although
some of them are very close to each other and may in fact be related (Table \ref{tab:bacillus_hotspots}).
These recombination hotspots are likely to represent regions of the genome
that are under positive selection, so that diversification via 
recombination provides a selective advantage which is more likely
to be observed in the data. For example, the hotspot around position 3.2\,Mbp
contains the \textit{bla2} gene which encodes a protein that provides broad-spectrum resistance against penicillins, cephalosporins, and carbapenems \citep{materon2003biochemical}. 
Likewise, gene GBAA3214 which encodes a beta-lactamase class C protein
is in a hotspot around position 2.95\,Mbp and is one of the genes subject to the highest positive selection in \textit{B. cereus} \citep{rasigadeGenesPositiveSelection2018}.
The lack of mutation hotspots suggests that in the \textit{B. cereus} group recombination is the driving evolutionary force of diversification as a result of its commensal lifestyle in a wide range of environments, which brings it in contact with many potential recombination partners providing a large reservoir of genetic diversity \citep{ceuppensDiversityBacillusCereus2013}.

\subsection{Application to \textit{Staphylococcus aureus}}

We reanalysed a previously described dataset of 110 genomes of \textit{Staphylococcus aureus} \citep{Everitt2014}. We first inferred a phylogenetic tree using RAxML-NG \citep{kozlov2019raxml} and applied ClonalFrameML \citep{didelot2015clonalframeml} to obtain an estimate of the clonal genealogy $\widehat{\mathcal{T}}_{110}$ and the mean recombination tract length, $\delta=454$\,bp. Simulations for training the species-specific NPE model were generated conditional on this genealogy using the independent priors
$\rho_s \sim \mathcal{U}(0,0.1)$, 
$\theta_s \sim \mathcal{U}(0,0.02)$ and 
$\ell \sim \operatorname{DiscreteUniform}\{20,\ldots,10{,}000\}$.
These prior ranges defined the parameter space considered in the \textit{S. aureus} analysis. For each simulation, summary statistics were calculated from the resulting gene sequence and supplied, together with gene length, as input to the NPE model. Generating the training data was again the most computationally demanding step. Each simulation required on average about 36 s of CPU time. Running 12 simulations in parallel on the same HPC compute node took approximately 58 h to obtain 50,000 simulations. Training the NPE model on these simulations for \textit{S. aureus} required approximately 27 minutes using the same CPU and packages.

\begin{figure}[H]
    \centering
    \includegraphics[width=15cm]{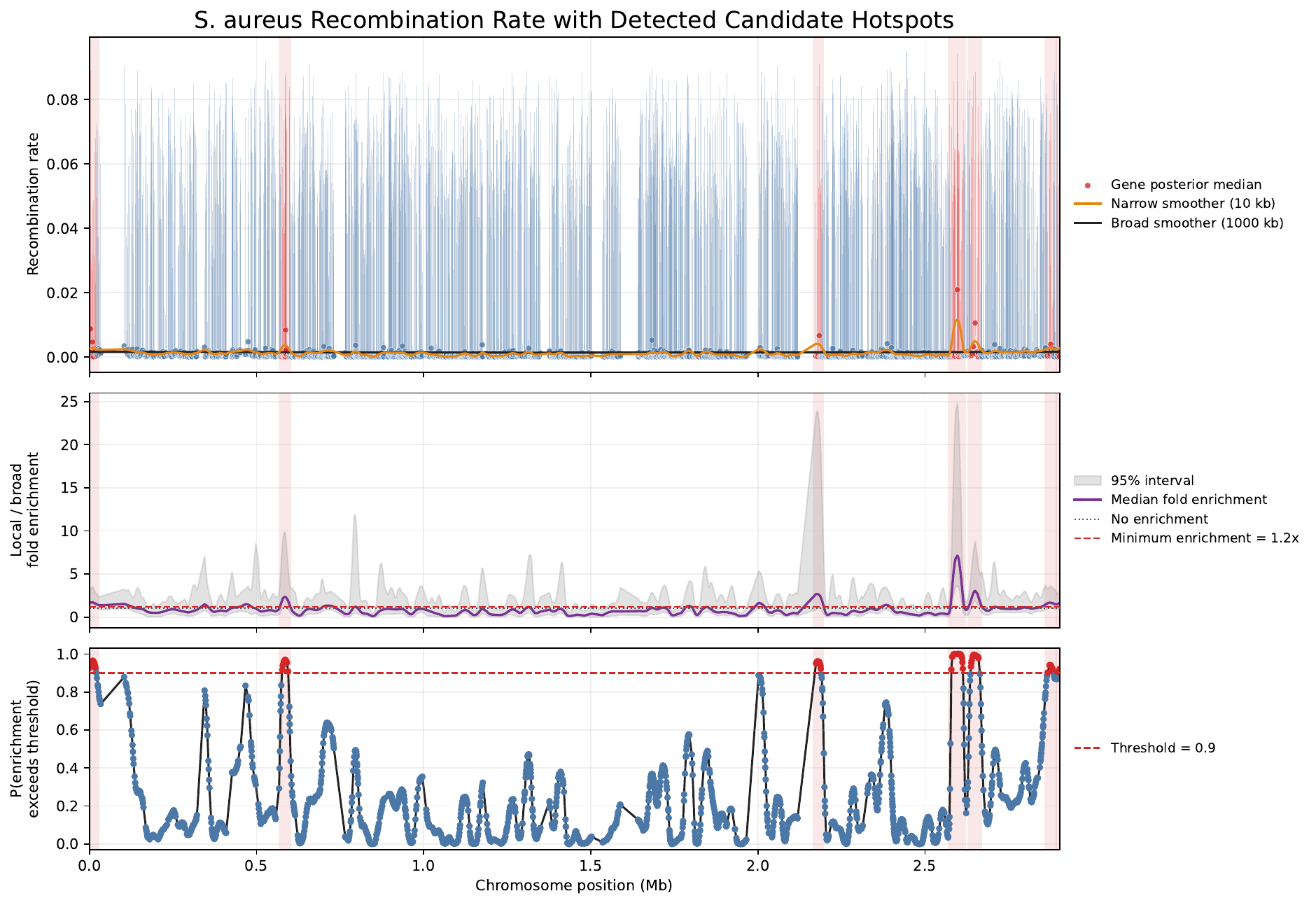}
    \caption{Gene-level posterior estimates of the local recombination parameter across the \textit{S. aureus} genome. Candidate recombination hotspots were identified using ppDoG with $h_{\mathrm{N}}=10{,}000\,\mathrm{bp}$, $h_{\mathrm{B}}=1{,}000{,}000\,\mathrm{bp}$, and $F_{\min}=1.2$.}
    \label{fig:staph_recomb}
\end{figure}

\begin{figure}[H]
    \centering
    \includegraphics[width=15cm]{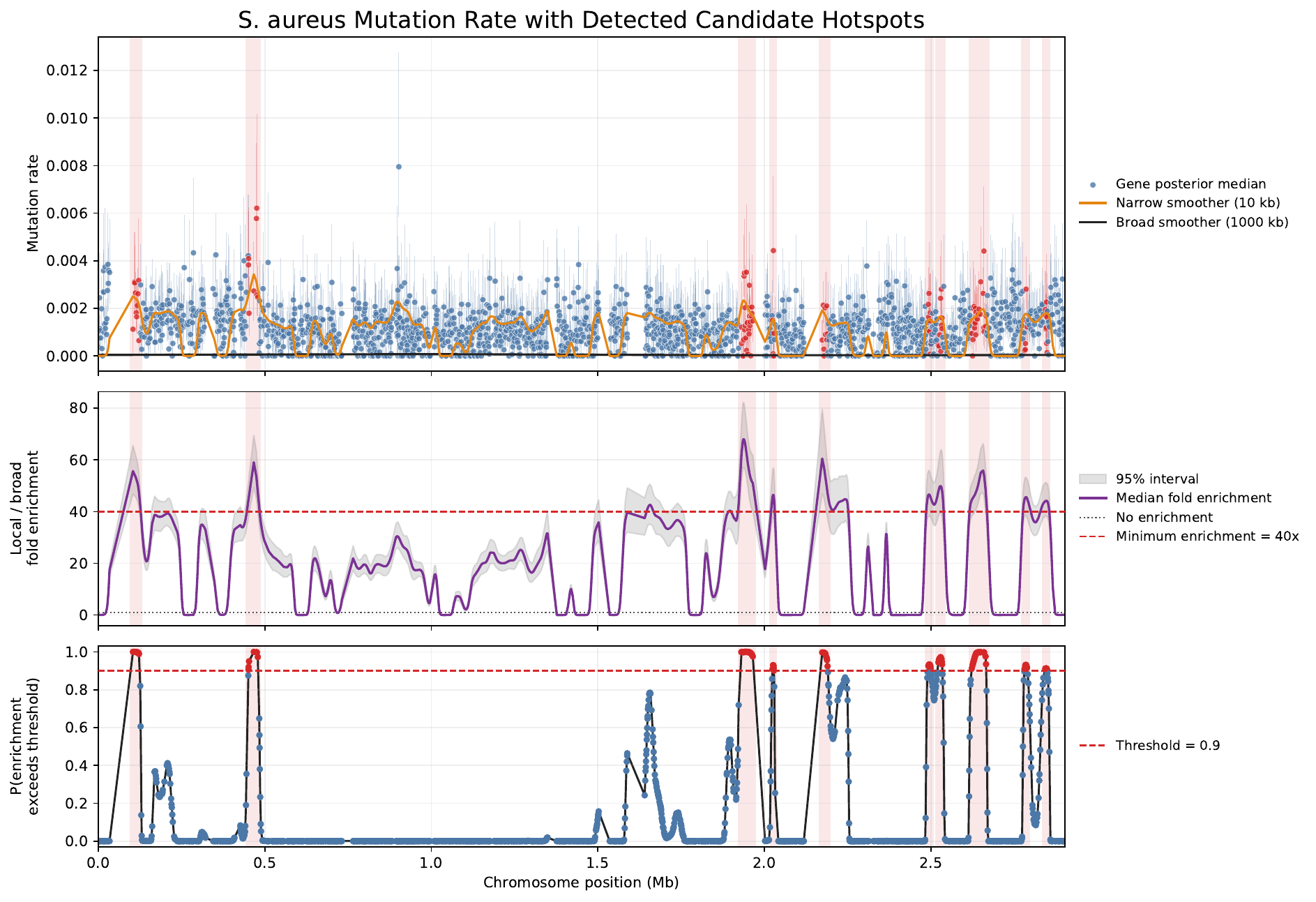}
    \caption{Gene-level posterior estimates of the local mutation parameter across the \textit{S. aureus} genome. Candidate mutation hotspots were identified using ppDoG with $h_{\mathrm{N}}=10{,}000\,\mathrm{bp}$, $h_{\mathrm{B}}=1{,}000{,}000\,\mathrm{bp}$, and $F_{\min}=40$.}
    \label{fig:staph_mutation}
\end{figure}

After training, the model was conditioned on the observed core genes of \textit{S. aureus} to obtain gene-specific posterior distributions for the local recombination and mutation parameters. The resulting estimates were analysed using ppDoG.  Figure~\ref{fig:staph_length} compares the inferred gene lengths with their observed values, providing reassurance on the model performance on this empirical data.
Figures~\ref{fig:staph_recomb} and \ref{fig:staph_mutation} present the ppDoG results for candidate hotspots of recombination and mutation.
It is noteworthy that the individual values for each analysed region have more uncertainty than in the previous analysis on the \textit{Bacillus cereus} group.
There are several reasons contributing to this difference. Firstly, each region is a biological gene, the length of which is rarely above 2{,}000\,bp, whereas the previous analysis was based on alignment blocks that could be up to 10{,}000\,bp.
Secondly, the previous analysis included several related species whereas the current one includes only \textit{S. aureus}. Finally, and partly as a result of the previous point, the mutation rate was overall higher in the previous analysis, with a mean around 0.04 whereas in \textit{S. aureus} the mean was about 30 times less. The uncertainty about the mutation and recombination rates of each gene is therefore quite high, but the ppDoG is still able to identify several hotspots of both mutation and recombination.

The \textit{S. aureus} analysis finds ten mutation and six recombination hotspots, listed in Table \ref{tab:staph_hotspots}. We qualitatively compare the results to \citep{Everitt2014}, which found an association between recombination (indicated by excess homoplasy) and insertion sites of mobile element. Five of the ten mutation hotspots are close to mobile element insertion sites documented in \citep{Everitt2014}. The ability of our new method to jointly infer mutation and recombination hotspots potentially indicates that some of the excess homoplasy described in \citep{Everitt2014} is attributable to high mutation rather than recombination. Two of the recombination hotspots are close to mobile element insertion sites from Everitt et al. (2014); the others are, as far as we know, previously undetected and warrant detailed analysis that is out of scope of this paper.

\subsection{Application to \textit{Escherichia coli}}

We reanalysed a dataset of 27 \textit{Escherichia coli} genomes, originally described by \citet{didelot2012impact} and since used by several other studies \citep{yahara2014efficient,Yahara2016,Medina2023}. 
We use the same 2675 alignment blocks and clonal genealogy as previously generated using progressiveMauve \citep{Darling2010} and ClonalFrame \citep{Didelot2007b} and the previous estimate of the mean recombination tract length, $\delta=542$\,bp \citep{didelot2012impact}. 
Simulations for training the NPE model were generated conditional on this genealogy using the independent priors
$\rho_s \sim \mathcal{U}(0,0.2)$, 
$\theta_s \sim \mathcal{U}(0,0.1)$ and 
$\ell \sim \operatorname{DiscreteUniform}\{400,\ldots,10{,}000\}$.
Figure~\ref{fig:ecoli_length} compares the inferred gene lengths with their observed values and does not reveal any problem with this inference. Each simulation required on average about 138 s of CPU time. Running 12 simulations in parallel on the same HPC compute node, 50,000 simulations took approximately 213 h of wall-clock time. Training the NPE model on these simulations for \textit{E. coli} required approximately 25 minutes using the same CPU and packages.

\begin{figure}[H]
    \centering
    \includegraphics[width=15cm]{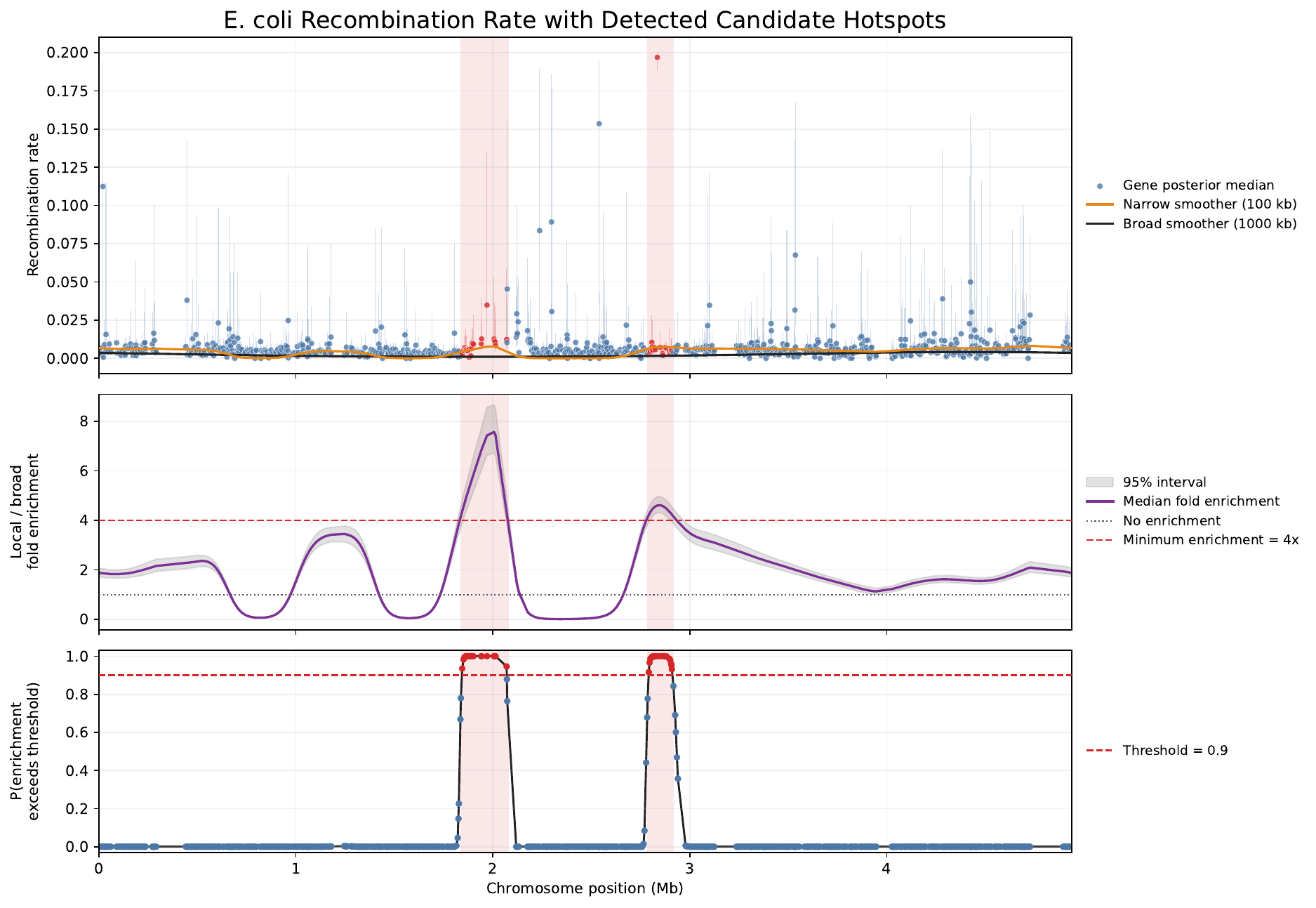}
    \caption{Gene-level posterior estimates of the local recombination parameter across the \textit{E. coli} genome. Candidate recombination hotspots were identified using ppDoG with $h_{\mathrm{N}}=100{,}000\,\mathrm{bp}$, $h_{\mathrm{B}}=1{,}000{,}000\,\mathrm{bp}$, and $F_{\min}=4$.}
    \label{fig:ecoli_recomb}
\end{figure}

\begin{figure}[H]
    \centering
    \includegraphics[width=15cm]{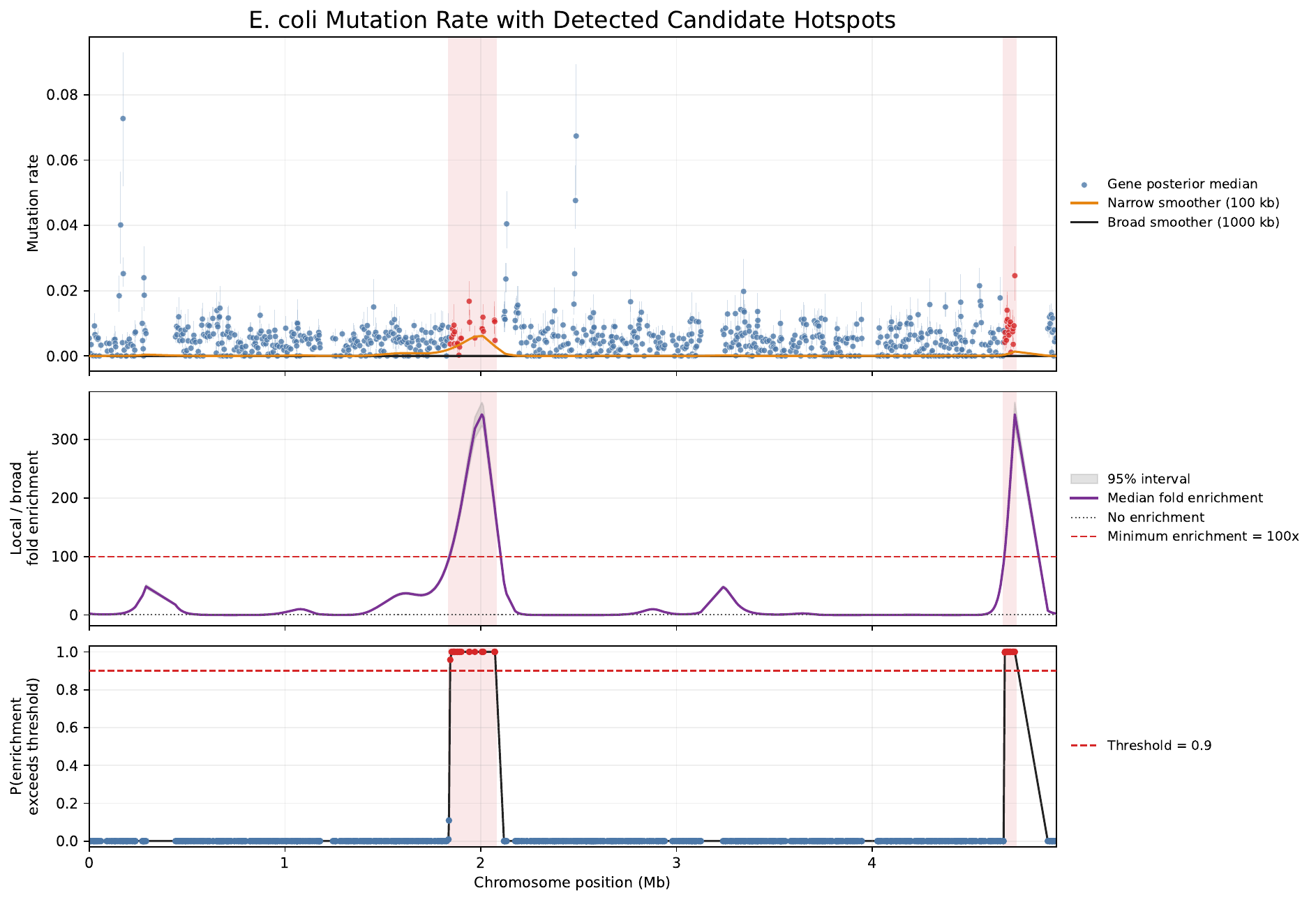}
    \caption{Gene-level posterior estimates of the local mutation parameter across the \textit{E. coli} genome. Candidate mutation hotspots were identified using ppDoG with $h_{\mathrm{N}}=100{,}000\,\mathrm{bp}$, $h_{\mathrm{B}}=1{,}000{,}000\,\mathrm{bp}$, and $F_{\min}=100.0$.}
    \label{fig:ecoli_mutation}
\end{figure}

Figures~\ref{fig:ecoli_recomb} and \ref{fig:ecoli_mutation} present the  analyses of the inferred recombination and mutation hotspots, using the coordinates of the reference genome \textit{E. coli} strain 536 \citep{brzuszkiewiczHowBecomeUropathogen2006}. Only two recombination hotspots and two mutation hotspots were found (Table \ref{tab:ecoli_hotspots}). The first recombination hotspot clearly coincides with the first mutation hotspot. This region includes the \textit{rfb} operon involved in the synthesis of the O antigen, and it has been previously reported as a recombination hotspot \citep{didelot2012impact,yahara2014efficient,Yahara2016}. However, the region around the \textit{fimA} gene towards the end of the genome has also been previously reported as a recombination hotspot \citep{didelot2012impact,yahara2014efficient,Yahara2016}, and here we found it to have an exceptionally high rate of mutation but not recombination. Since the previous methods used to detect recombination hotspot assume a constant rate of mutation throughout the genome, this suggests that the mutation hotspot was previously misclassified as a recombination hotspot. This result highlights the importance in our methodology of considering both the mutation and recombination rates as variable along the genome.

\section{DISCUSSION}

We have described a new statistical method to analyse and detect genomic variations in the mutation and recombination rates of bacteria. By testing the method on simulated datasets, in which the true parameter values are known, we showed that our method is able to recover the correct values. Our methodology is based on neural posterior estimation \citep{papamakarios2016fast, lueckmann2017flexible, greenberg2019automatic} which allowed us to use a much wider set of summary statistics as input compared to a previous comparable method based on approximate Bayesian computation \citep{ansari2014inference}. Jointly accounting for variations in the mutation rates also makes our method more robust than previous attempts at detecting recombination hotspots \citep{yahara2014efficient,Yahara2016}. 
We presented three examples of application to real datasets which showed that the method can perform well in a wide range of scenarios, and was able to detect previously described hotspots as well as make new discoveries relative to previous analyses of the same datasets \citep{didelot2010inference,didelot2012impact,ansari2014inference,Everitt2014}.
Most importantly, our method is implemented into open source software which should be useful for many scientific researchers working on bacterial genomics.

A limitation of our approach is its dependence on the assumptions encoded by the simulator used for training. Although NPE can provide rapid and well-calibrated inference for observations represented by the training distribution, it does not correct for discrepancies between the simulator and the true data-generating process. The simulator we used is based on the ClonalOrigin model \citep{didelot2010inference} which is especially convenient since it allows us to simulate the recombination process conditional on the clonal genealogy, so that several genomic regions can be simulated independently. The ClonalOrigin model is an approximation of the ancestral recombination graph model with gene conversion \citep{Wiuf2000} and it could therefore be interesting for future work to investigate the use of the full ARG model as a simulator, or another approximation such as the bacterial sequential Markov coalescent 
model \citep{Maio2017a}.

The present analysis primarily resolves variation in recombination and mutation parameters at the gene level, with a posterior-propagated Difference-of-Gaussians method applied as a postprocessing step in order to detect hotspots at a wider genomic level. Future work could build on this framework by developing more comprehensive simulation models that impose different restrictions on recombination events. The simulation framework could also be extended, for example, to models with biased recombination \citep{ansari2014inference}, natural selection, changes in population size, or migrations between structured populations, all of which could affect the recombination process \citep{Vaughan2017}. Training NPE models on simulations spanning these alternative mechanisms would make it possible to estimate additional evolutionary parameters, evaluate sensitivity to competing biological assumptions, and reduce bias when analysing empirical genomes. More flexible simulators, combined with validation across multiple data-generating models, could ultimately improve the robustness and biological interpretability of simulation-based inference.

\printbibliography
\end{refsection}

\clearpage
\setcounter{figure}{0}
\setcounter{table}{0}
\setcounter{section}{0}
\setcounter{equation}{0}
\makeatletter 
\renewcommand{\thefigure}{S\@arabic\c@figure} 
\renewcommand{\thetable}{S\@arabic\c@table} 
\renewcommand{\thesection}{S\@arabic\c@section} 
\renewcommand{\theequation}{S\@arabic\c@equation} 
\makeatother

\begin{refsection}
\setcounter{page}{1}
{\Huge\textbf{Supplementary Material}}

\section{ClonalOrigin simulation algorithm}
\label{appendix:pseudo_code}

This section gives the implementation-level procedure used to simulate a sequence alignment under the ClonalOrigin model described in the main text. The simulation begins with a fixed clonal genealogy $\mathcal{T}$ and its total branch length $T$. Recombination events covering the first site are generated from Equation~\eqref{eq:R_1}, after which a departure point and an arrival point are sampled for each event using Equations~\eqref{eq:p_b} and \eqref{eq:p_a}. For every subsequent site, active recombination events are retained according to Equation~\eqref{eq:R_s}, and newly initiated events are added according to Equation~\eqref{eq:R_n}. The active events are then superimposed on the clonal genealogy to construct the local genealogy at that site. Finally, mutations are generated on each local genealogy according to Equation~\eqref{eq:M_i}, and their locations are sampled along its branches. Algorithm~\ref{pseudo_code:ClonalOrigin} summarises this procedure.

\begin{algorithm}
\caption{Sequential simulation under the ClonalOrigin model}
\label{pseudo_code:ClonalOrigin}
\begin{algorithmic}[1]
\REQUIRE Number of sampled genomes $n$; recombination parameter $\rho_s$; mutation parameter $\theta_s$; mean recombination-tract length $\delta$; sequence length $\ell$; clonal genealogy $\mathcal{T}$.
\STATE Compute the total branch length $T$ of $\mathcal{T}$.
\STATE Draw the number of recombination events affecting the first site, $R_1$, from Equation~\eqref{eq:R_1}.
\FOR{$e=1,\ldots,R_1$}
    \STATE Draw the departure and arrival points, $b_e^1$ and $a_e^1$, using Equations~\eqref{eq:p_b} and \eqref{eq:p_a}.
\ENDFOR
\STATE Construct the local genealogy at the first site, denoted by $\mathcal{T}_1$.
\FOR{$i=1,\ldots,\ell-1$}
    \STATE Draw the number of surviving events $R_{i+1}^{\mathrm{S}}$ from Equation~\eqref{eq:R_s}.
    \STATE Draw the number of newly initiated events $R_{i+1}^{\mathrm{N}}$ from Equation~\eqref{eq:R_n}.
    \STATE Set $R_{i+1}=R_{i+1}^{\mathrm{S}}+R_{i+1}^{\mathrm{N}}$.
    \FOR{each newly initiated event, $e$}
        \STATE Draw its departure and arrival points, $b_e^{i+1}$ and $a_e^{i+1}$, using Equations~\eqref{eq:p_b} and \eqref{eq:p_a}.
    \ENDFOR
    \STATE Construct the local genealogy $\mathcal{T}_{i+1}$ from $\mathcal{T}$ and the recombination events active at site $i+1$.
\ENDFOR
\FOR{$i=1,\ldots,\ell$}
    \STATE Draw $M_i$ from Equation~\eqref{eq:M_i}.
    \STATE Place the $M_i$ mutation events along the branches of $\mathcal{T}_i$ and record the resulting states at the leaves.
\ENDFOR
\RETURN The simulated sequence alignment of $n$ sequences and length $\ell$.
\end{algorithmic}
\end{algorithm}

\section{Summary statistics}
\label{appendix:summary_stats}

For each simulated or observed gene alignment of $n$ sequences and length $\ell$, we computed the summary statistics defined below. Throughout this section, $S$ denotes the number of segregating sites in the alignment.

\textbf{Watterson's estimator} is defined as \citep{Watterson1975}:

\begin{align}
    \hat{\theta}_W = \frac{S}{a_n} \quad\mathrm{with~} a_n = \sum_{j=1}^{n-1} \frac{1}{j},
\end{align}
where $a_n$ is proportional to the expected total branch length of the coalescent tree of $n$ sequences.

\textbf{Tajima's estimator} is defined as
\citep{Tajima1983}:

\begin{align}
    \hat{\pi}_T = \frac{2}{n(n-1)} \sum_{i < j} \pi_{ij},
\end{align}
where $\pi_{ij}$ is the number of sites at which sequences $i$ and $j$ differ.

\textbf{Tajima's D} is defined as \citep{tajima1989statistical}:

\begin{align}
    D_T = \frac{\hat{\pi}_T - \hat{\theta}_W}{\sqrt{e_1S + e_2S(S - 1)}},
\end{align}
where
\begin{align}
    e_1 &= \frac{n + 1}{3a_n(n - 1)} - \frac{1}{a_n^2},\\
    e_2 &= \frac{1}{a_n^2 + b_n} \left( \frac{2(n^2 + n + 3)}{9n(n - 1)} - \frac{n + 2}{na_n} + \frac{b_n}{a_n^2} \right),\\
    b_n &= \sum_{j=1}^{n-1} \frac{1}{j^2}.
\end{align}
Thus $D_T$ is the normalised difference between $\hat{\pi}_T$ and $\hat{\theta}_W$. Under the standard neutral coalescent, $D_T$ has mean approximately zero and variance approximately one, and values satisfying $\lvert D_T\rvert\leq2$ are usually taken to indicate that the data are consistent with the standard coalescent model.

The \textbf{proportion of segregating sites} is simply:

\begin{align}
    p_S = \frac{S}{\ell}.
\end{align}

The \textbf{homoplasy index} measures the amount of homoplasy in the data:
\begin{align}
    HI = 1 - \frac{M}{K},
\end{align}
where $M$ is the minimum number of changes required, and $K$ is the actual number of changes observed.

\textbf{Wakeley's estimator} is defined as \citep{Wakeley1997}:

\begin{align}
    \hat{\pi}_W = \frac{2}{n(n-1)} \sum_{i < j} (\pi_{ij} - \hat{\pi}_T)^2,
\end{align}
which measures the variance of the pairwise differences around their mean.

\textbf{Linkage disequilibrium (LD)} between two segregating sites $i$ and $j$ is captured by several statistics \citep{Lewontin1964}. Let $p_{11}$ denote the frequency of sequences carrying allele 1 at both sites and $p_{01}$ the frequency of sequences carrying allele 0 at site $i$ and allele 1 at site $j$, with $p_{00}$ and $p_{10}$ defined similarly. The allele frequencies at sites $i$ and $j$ are then $p_1 = p_{10} + p_{11}$ and $q_1 = p_{01} + p_{11}$, respectively, with $p_0=1-p_1$ and $q_0=1-q_1$. The following LD statistics are defined:
\begin{align}
    D &= p_{00}p_{11} - p_{01}p_{10} = p_{00} - p_0q_0,\\
    D' &=
        \begin{cases}
        \frac{D}{\min(p_0q_1, p_1q_0)} & \text{if } D > 0, \\
        \frac{-D}{\min(p_0q_0, p_1q_1)} & \text{if } D < 0,
        \end{cases}\\
    r^2 &= \frac{D^2}{p_0p_1q_0q_1}.
\end{align}
The statistic $D$ measures the departure from statistical independence between the alleles at sites $i$ and $j$, $D'$ rescales $D$ by its maximum possible magnitude given the allele frequencies, and $r^2$ is the squared correlation between the alleles at the two sites.

\textbf{Four-gamete (G4) test.} A pair of segregating sites $i$ and $j$ fails the four-gamete test if all four possible two-site haplotypes ($00$, $01$, $10$ and $11$) are observed in the sample. A failed test indicates that at least one recombination event (or a recurrent mutation) has occurred between the two sites \citep{Hudson1985}. We write $G_{ij}=1$ if the pair $(i,j)$ fails the test and $G_{ij}=0$ otherwise.

\textbf{Distance-stratified LD and G4 statistics.} Because recombination breaks down associations more effectively between distant sites, we summarised the pairwise statistics separately for close and distant pairs of sites. Let $d_{ij}$ denote the physical distance between segregating sites $i$ and $j$, and partition the set $\mathcal{P}$ of all pairs of segregating sites into
\begin{equation}
    \mathcal{P}_{\mathrm{near}}=\{(i,j)\in\mathcal{P}: d_{ij}<\ell/2\}
    \quad\text{and}\quad
    \mathcal{P}_{\mathrm{far}}=\{(i,j)\in\mathcal{P}: d_{ij}\geq\ell/2\},
\end{equation}
where $\ell$ is the gene length. For each class $k\in\{\mathrm{near},\mathrm{far}\}$ and each pairwise quantity $X_{ij}\in\{D_{ij},D'_{ij},r^2_{ij},G_{ij}\}$, we computed an exact statistic and a proportional statistic,
\begin{equation}
    X_k^{\mathrm{exact}}=\sum_{(i,j)\in\mathcal{P}_k}X_{ij}
    \qquad\text{and}\qquad
    X_k^{\mathrm{prop}}=\frac{1}{\lvert\mathcal{P}_k\rvert}\sum_{(i,j)\in\mathcal{P}_k}X_{ij}.
\end{equation}
This gives eight exact and eight proportional statistics; these are the ``exact LD and G4 features'' and ``proportional LD and G4 features'' referred to in Table~\ref{tab:npe_tuning_space}.

\textbf{Kelly's statistics} are defined as 
\citep{kelly1997test}:
\begin{align}
    Z_{nS} = \frac{2}{S (S - 1)} \sum_{i=1}^{S - 1} \sum_{j = i+1}^{S} r^2_{ij}.
\end{align}

\textbf{Hudson's estimator $R_M$}
estimates the theoretically minimal recombination numbers for the whole sequence \citep{Hudson1985}. It finds the pairs failing a G4 test and find the minimal number of recombination that can explain these pairs.

\textbf{Wall's statistics} are defined as follows \citep{wall1999recombination}.
We say that two segregating sites are congruent if and only if their corresponding partitions are identical. Let $B'$ be the number of pairs of adjacent segregating sites that are congruent. And let $A$ be the set of all distinct partitions induced by congruent pairs of segregating sites. Wall's statistics are then
\begin{align}
    B &= \frac{B'}{S - 1},\\
    Q &= \frac{B' + |A|}{S}.
\end{align}

\textbf{Regression coefficients}
of the LD measure can also be used.
For each pair of segregating sites $(i,j)$, we then fit an exponential regression model of the form
\begin{align}
r_{ij}^2 = a \exp(-b d_{ij}) + c + \varepsilon_{ij},
\end{align}
for all pairs $(i,j) \in \mathcal{P}$, where $\mathcal{P}$ denotes the set of all pairs of segregating sites and $\varepsilon_{ij}$ is the residual error term. The estimated regression coefficients, $\hat{a}$, $\hat{b}$, and $\hat{c}$, are then used as summary statistics.

\section{Neural posterior estimation (NPE)}
\label{appendix:npe}

This section gives the complete NPE procedure used for every analysis in this paper. It combines the posterior-estimation objective of Section~\ref{sec:sbi} with the length-preserving compression network of Section~\ref{sec:compression_network}; the full procedure is summarised in Algorithm~\ref{pseudo_code:npe}.

\begin{algorithm}
\caption{NPE-C for local genomic parameters}
\label{pseudo_code:npe}
\begin{algorithmic}[1]
\REQUIRE Prior $\pi(\params)$; simulator $p(z\mid\params)$ (Algorithm~\ref{pseudo_code:ClonalOrigin} followed by the summary statistics of Supplementary Text~\ref{appendix:summary_stats}); number of simulations $N$; observed genes $z_{\mathrm{obs},1},\ldots,z_{\mathrm{obs},G}$; number of posterior draws $M$.
\FOR{$i=1,\ldots,N$}
    \STATE Sample parameters $\params_i=(\rho_{s,i},\theta_{s,i},\ell_i)\sim\pi(\params)$.
    \STATE Simulate an alignment of length $\ell_i$ and compute $z_i=(s_{1,i},\ldots,s_{d,i},\ell_i)\sim p(z\mid\params_i)$.
\ENDFOR
\STATE Split $\{(\params_i,z_i)\}_{i=1}^{N}$ into training and validation sets.
\REPEAT
    \STATE Update $(\phi,\psi_{\mathrm{e}})$ by a gradient step on the loss, evaluated on a minibatch of training pairs.
\UNTIL{the validation loss has not improved for a fixed number of epochs.}
\STATE Denote the trained parameters by $(\widehat{\phi},\widehat{\psi}_{\mathrm{e}})$.
\FOR{$g=1,\ldots,G$}
    \STATE Compute the embedding $x_g=e_{\widehat{\psi}_{\mathrm{e}}}(z_{\mathrm{obs},g})$.
    \STATE Draw $\params_g^{(m)}=f_{\widehat{\phi}}(u^{(m)};x_g)$ with $u^{(m)}\sim\mathcal{N}(0,I_3)$ for $m=1,\ldots,M$, discarding draws outside the prior support.
\ENDFOR
\RETURN Posterior draws $\{\params_g^{(m)}\}$ from $\widehat{\pi}(\params\mid z_{\mathrm{obs},g})=q_{\widehat{\phi}}\bigl(\params\mid e_{\widehat{\psi}_{\mathrm{e}}}(z_{\mathrm{obs},g})\bigr)$, $g=1,\ldots,G$.
\end{algorithmic}
\end{algorithm}

\section{Neural likelihood estimation (NLE)}
\label{appendix:nle}

Neural likelihood estimation (NLE) \citep{papamakarios2019sequential} was included only as a benchmark. It uses the same simulated pairs $(\params_i,z_i)$ as NPE but fits a conditional density estimator $q_{\phi}(z\mid\params)$ to approximate the likelihood $p(z\mid\params)$. The estimator is trained by minimising the negative log-likelihood of the simulated observations,
\begin{equation}
    \mathcal{L}_{\mathrm{NLE}}(\phi)
    =
    -\frac{1}{N}\sum_{i=1}^{N}\log q_{\phi}(z_i\mid\params_i).
\end{equation}
We used a conditional neural spline flow for $q_{\phi}(z\mid\params)$. For an observed segment, the learned likelihood is combined with the prior to define
\begin{equation}
    \widehat{\pi}(\params\mid z_{\mathrm{obs}})
    \propto q_{\widehat{\phi}}(z_{\mathrm{obs}}\mid\params)\,\pi(\params).
\end{equation}
Unlike NPE, NLE therefore requires a separate MCMC run for each observed segment to sample from this posterior.

\section{Neural ratio estimation (NRE)}
\label{appendix:nre}

Neural ratio estimation (NRE) \citep{hermans2020likelihood} was also used only for benchmarking. Rather than estimating a posterior or likelihood density, NRE learns the likelihood-to-evidence ratio
\begin{equation}
    r(\params,z)
    =
    \frac{p(z\mid\params)}{p(z)}
    =
    \frac{\pi(\params\mid z)}{\pi(\params)}.
\end{equation}
A classifier $d_{\phi}(\params,z)\in(0,1)$ is trained to distinguish joint samples $(\params,z)\sim p(\params,z)$ from independently paired samples $(\params,z)\sim\pi(\params)p(z)$. For the Bayes-optimal classifier, the odds $d_{\phi}/(1-d_{\phi})$ equal $r(\params,z)$, so the trained classifier provides the ratio estimate $\widehat{r}_{\widehat{\phi}}(\params,z)=d_{\widehat{\phi}}(\params,z)/\{1-d_{\widehat{\phi}}(\params,z)\}$. In our benchmark, we used a residual neural network classifier and generated the independent pairs by shuffling observations within training batches.

For an observed segment, the estimated ratio is combined with the prior:
\begin{equation}
    \widehat{\pi}(\params\mid z_{\mathrm{obs}})
    \propto
    \widehat{r}_{\phi}(\params,z_{\mathrm{obs}})\,\pi(\params).
\end{equation}
As with NLE, this posterior is available only up to a normalising constant and requires MCMC for each observed segment. These additional sampling costs, together with the benchmark results in Supplementary Text~\ref{appendix:benchmark}, motivate our use of NPE for the main analysis.

\section{Circular posterior-propagated Difference-of-Gaussians hotspot detection}
\label{appendix:ppdog}

This section gives the full definition of the ppDoG detector summarised in Section~\ref{sec:ppdog}. Let a circular chromosome of length $L$ contain $G$ genes at representative genomic coordinates $x_1,\ldots,x_G\in[0,L)$. For gene $i$, let $\lambda_i>0$ denote either its recombination parameter $\rho_s$ or its mutation parameter $\theta_s$; the two rates are analysed separately. The trained NPE model provides $M$ posterior draws $\lambda_i^{(1)},\ldots,\lambda_i^{(M)}$ from the corresponding marginal of $q_{\widehat{\phi}}\bigl(\params\mid e_{\widehat{\psi}_{\mathrm{e}}}(z_{\mathrm{obs},i})\bigr)$ (Algorithm~\ref{pseudo_code:npe}), where $z_{\mathrm{obs},i}$ is the observed input for gene $i$. We work with the log rates
\begin{equation}
    \Lambda_i=\log\lambda_i,
    \qquad
    \Lambda_i^{(m)}=\log\lambda_i^{(m)},
    \quad m=1,\ldots,M.
    \label{eq:appendix_ppdog_log_rate}
\end{equation} Filtering is performed on the natural-log scale because the rates are positive and may span several orders of magnitude.

To respect chromosome circularity, the distance between genes $i$ and $j$ is defined as
\begin{equation}
    d_{ij}
    =
    \min\left\{
        \lvert x_i-x_j\rvert,
        L-\lvert x_i-x_j\rvert
    \right\}.
    \label{eq:appendix_circular_distance}
\end{equation}
Genes near coordinate zero are therefore treated as neighbours of genes near coordinate $L$.

Posterior uncertainty is incorporated through the variance of each gene's log-rate posterior,
\begin{equation}
    v_j
    =
    \operatorname{Var}\!\left(
        \Lambda_j\mid z_{\mathrm{obs},j}
    \right),
\end{equation}
which is estimated by the sample variance of $\Lambda_j^{(1)},\ldots,\Lambda_j^{(M)}$. A stabilised precision weight is assigned as
\begin{equation}
    p_j
    =
    \frac{1}{v_j+\tau^2},
    \label{eq:appendix_ppdog_precision}
\end{equation}
where $\tau^2$ represents residual log-rate variability and prevents extremely concentrated gene-level posteriors from dominating the smoother. For bandwidth $h$, the unnormalised and normalised weights are
\begin{align}
    w_{ij}(h)
    &=
    p_j
    \exp\left(
        -\frac{d_{ij}^2}{2h^2}
    \right),\\
    W_{ij}(h)
    &=
    \frac{w_{ij}(h)}
         {\sum_{k=1}^{G}w_{ik}(h)},
    \qquad
    \sum_{j=1}^{G}W_{ij}(h)=1.
    \label{eq:appendix_ppdog_weights}
\end{align}
Thus, a source gene contributes more strongly when it is closer to the target coordinate and its posterior log rate is more precisely estimated.

Let $h_{\mathrm{N}}$ and $h_{\mathrm{B}}$ denote the narrow and broad bandwidths, with $0<h_{\mathrm{N}}<h_{\mathrm{B}}$. For posterior draw $m$, the narrow-scale local signal and broad-scale background at gene $i$ are
\begin{align}
    N_i^{(m)}
    &=
    \sum_{j=1}^{G}
    W_{ij}(h_{\mathrm{N}})
    \Lambda_j^{(m)},\\
    B_i^{(m)}
    &=
    \sum_{j=1}^{G}
    W_{ij}(h_{\mathrm{B}})
    \Lambda_j^{(m)}.
    \label{eq:appendix_ppdog_smoothers}
\end{align}
The posterior-propagated Difference-of-Gaussians score is
\begin{equation}
    D_i^{(m)}
    =
    N_i^{(m)}-B_i^{(m)}.
    \label{eq:appendix_ppdog_score}
\end{equation}
The two-scale contrast acts as a spatial band-pass filter \citep{lindeberg1998feature,lowe2004distinctive}. Positive values indicate local enrichment relative to the regional background, negative values indicate local depletion, and zero indicates no contrast between the two scales. Because the score is a difference of log rates,
\begin{equation}
    F_i^{(m)}
    =
    \exp\!\left(D_i^{(m)}\right)
    =
    \frac{\exp\!\left(N_i^{(m)}\right)}
         {\exp\!\left(B_i^{(m)}\right)}
    \label{eq:appendix_ppdog_fold}
\end{equation}
is the narrow-to-broad fold enrichment.

For a specified minimum enrichment $F_{\min}>1$, the posterior hotspot probability is
\begin{equation}
    P_i^{\mathrm{hot}}
    =
    \Pr\!\left(
        D_i>\log F_{\min}
        \mid\mathbf{Z}_{\mathrm{obs}}
    \right),
    \qquad
    \mathbf{Z}_{\mathrm{obs}}
    =
    (z_{\mathrm{obs},1},\ldots,z_{\mathrm{obs},G}),
    \label{eq:appendix_ppdog_probability}
\end{equation}
where $D_i=N_i-B_i$ is the score computed from the log rates $(\Lambda_1,\ldots,\Lambda_G)$.
This probability is conditioned on the chromosome-wide set of observations because the local and background smoothers combine rates from multiple genes. Using $M$ joint posterior draws, it is estimated by
\begin{equation}
    \widehat{P}_i^{\mathrm{hot}}
    =
    \frac{1}{M}
    \sum_{m=1}^{M}
    \mathbb{I}\!\left[
        D_i^{(m)}>\log F_{\min}
    \right].
    \label{eq:appendix_ppdog_probability_hat}
\end{equation}
As stated in the main text, genes with $\widehat{P}_i^{\mathrm{hot}}\geq0.9$ are classified as candidate hotspot genes. Runs of adjacent candidate genes along the circular chromosome are then merged into candidate hotspot regions, which are reported in Tables~\ref{tab:bacillus_hotspots}--\ref{tab:ecoli_hotspots}.

\textbf{Interpretation of the bandwidths.} The two bandwidths set the genomic scales to which the detector responds. The narrow bandwidth $h_{\mathrm{N}}$ should be comparable to the expected width of a localised enrichment, and should span at least several neighbouring genes. If $h_{\mathrm{N}}$ is much smaller than the spacing between genes, $N_i^{(m)}$ is essentially determined by gene $i$ alone and the score becomes sensitive to gene-level noise. If $h_{\mathrm{N}}$ is much wider than the enrichment, the peak is averaged with its surroundings. The broad bandwidth $h_{\mathrm{B}}$ should be substantially larger than $h_{\mathrm{N}}$, so that $B_i^{(m)}$ follows regional variation without following the local peak, yet small enough relative to the chromosome length that it still tracks large-scale variation along the chromosome rather than reducing to a single chromosome-wide average value. Variation on scales much broader than $h_{\mathrm{B}}$ is absorbed into the background and removed, features much narrower than the spacing between observed genes cannot be resolved reliably. The resulting calls should therefore be interpreted as scale-dependent evidence for relative local enrichment, rather than as direct evidence of a specific hotspot mechanism.

\textbf{Choice of $\tau$, $h_{\mathrm{N}}$, $h_{\mathrm{B}}$ and $F_{\min}$ in this paper.}
There is no universally optimal choice of these settings, because the appropriate values depend on gene density, chromosome length and the scale of enrichment of interest. We therefore adopted the following practical strategy, and the resulting values are listed in Table~\ref{tab:ppdog_settings}. First, the stabilising constant was fixed at $\tau=0.2$ in all applications. Second, the broad bandwidth was fixed at $h_{\mathrm{B}}=1{,}000{,}000$\,bp ($1$\,Mbp) in all applications. This value is large compared with individual genes and with the hotspot regions of interest, but corresponds to only a fraction of the chromosomes analysed here (approximately $2.9$--$5.2$\,Mbp), so that the background still follows regional variation. Third, with $h_{\mathrm{B}}$ fixed, $h_{\mathrm{N}}$ and $F_{\min}$ were adjusted for each dataset and rate type so that the narrow smoother provided a clear local signal distinct from the background, and the candidate regions were neither so wide and few that distinct peaks were merged, nor so narrow and numerous that isolated gene-level fluctuations were called. The hotspot probability threshold of $0.9$ was kept fixed throughout. Because $F_{\min}$ is a threshold on enrichment relative to the background, its appropriate value depends on the range of local-to-background contrasts in each dataset; the large values used for the mutation rate in \textit{S. aureus} and \textit{E. coli} reflect the much stronger mutation-rate contrasts observed in these datasets (see the peak fold enrichments in Tables~\ref{tab:staph_hotspots} and \ref{tab:ecoli_hotspots}). When applying the method to new data, the same strategy can be followed, and we recommend checking that the main candidate regions remain stable under moderate changes of $h_{\mathrm{N}}$ and $F_{\min}$.

\begin{table}[htbp]
    \centering
    \caption{ppDoG settings used in the real-data applications. In all cases, $\tau=0.2$ and genes with $\widehat{P}_i^{\mathrm{hot}}\geq0.9$ were classified as candidate hotspot genes.}
    \label{tab:ppdog_settings}
    \begin{tabular}{llccc}
        \hline
        \textbf{Dataset} & \textbf{Rate} & $h_{\mathrm{N}}$ (bp) & $h_{\mathrm{B}}$ (bp) & $F_{\min}$ \\
        \hline
        \textit{B. cereus} group & Recombination & $10{,}000$  & $1{,}000{,}000$ & $1.5$ \\
        \textit{B. cereus} group & Mutation      & $10{,}000$  & $1{,}000{,}000$ & $1.5$ \\
        \textit{S. aureus}       & Recombination & $10{,}000$  & $1{,}000{,}000$ & $1.2$ \\
        \textit{S. aureus}       & Mutation      & $10{,}000$  & $1{,}000{,}000$ & $40$ \\
        \textit{E. coli}         & Recombination & $100{,}000$ & $1{,}000{,}000$ & $4$ \\
        \textit{E. coli}         & Mutation      & $100{,}000$ & $1{,}000{,}000$ & $100$ \\
        \hline
    \end{tabular}
\end{table}

\section{Comparison between NPE, NLE and NRE}
\label{appendix:benchmark}

\begin{figure}[H]
\begin{center}
\includegraphics[width=10cm]{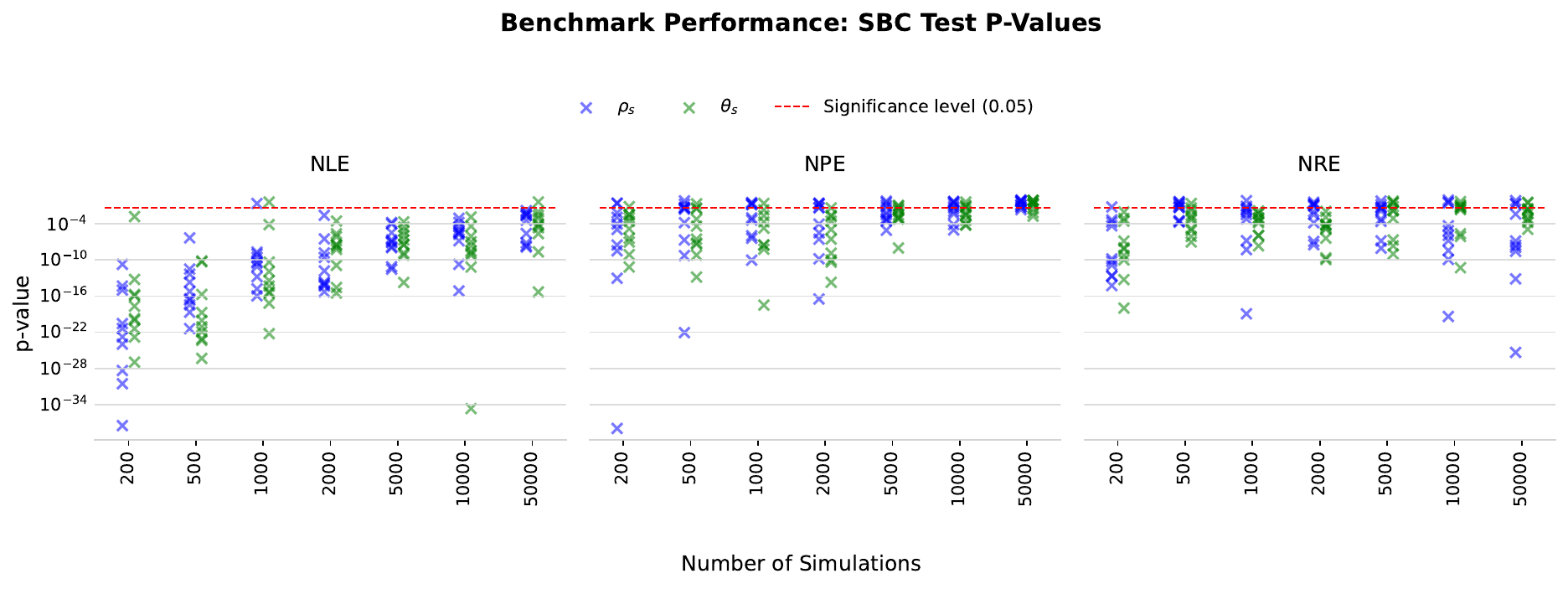}
\end{center}
\caption{Comparison of NPE, NLE, and NRE across training simulation budgets using the $p$-values from one-sample Kolmogorov--Smirnov tests applied to the normalised SBC ranks.
\label{fig:benchmark_sbc_p}}
\end{figure}

\begin{figure}[H]
\begin{center}
\includegraphics[width=10cm]{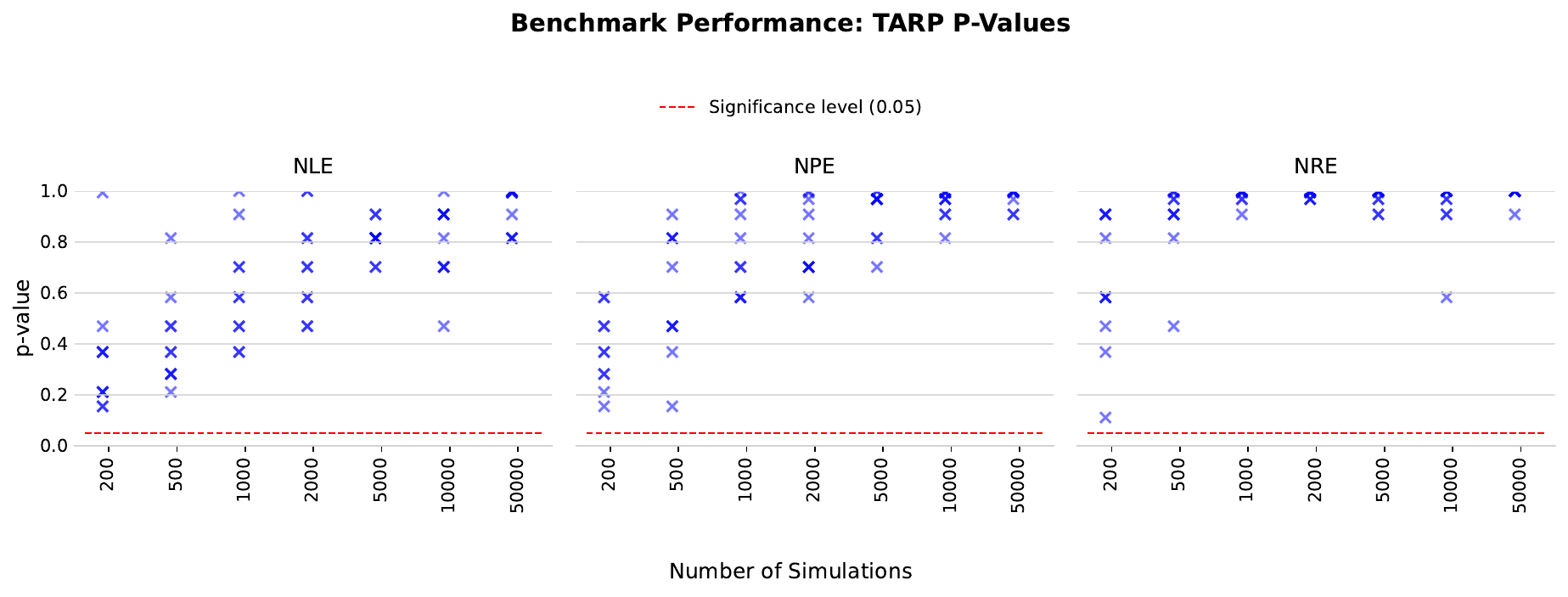}
\end{center}
\caption{Comparison of NPE, NLE, and NRE across training simulation budgets using the $p$-values from one-sample Kolmogorov--Smirnov tests applied to the normalised TARP ranks.
\label{fig:benchmark_tarp_p}}
\end{figure}

\begin{figure}[H]
\begin{center}
\includegraphics[width=10cm]{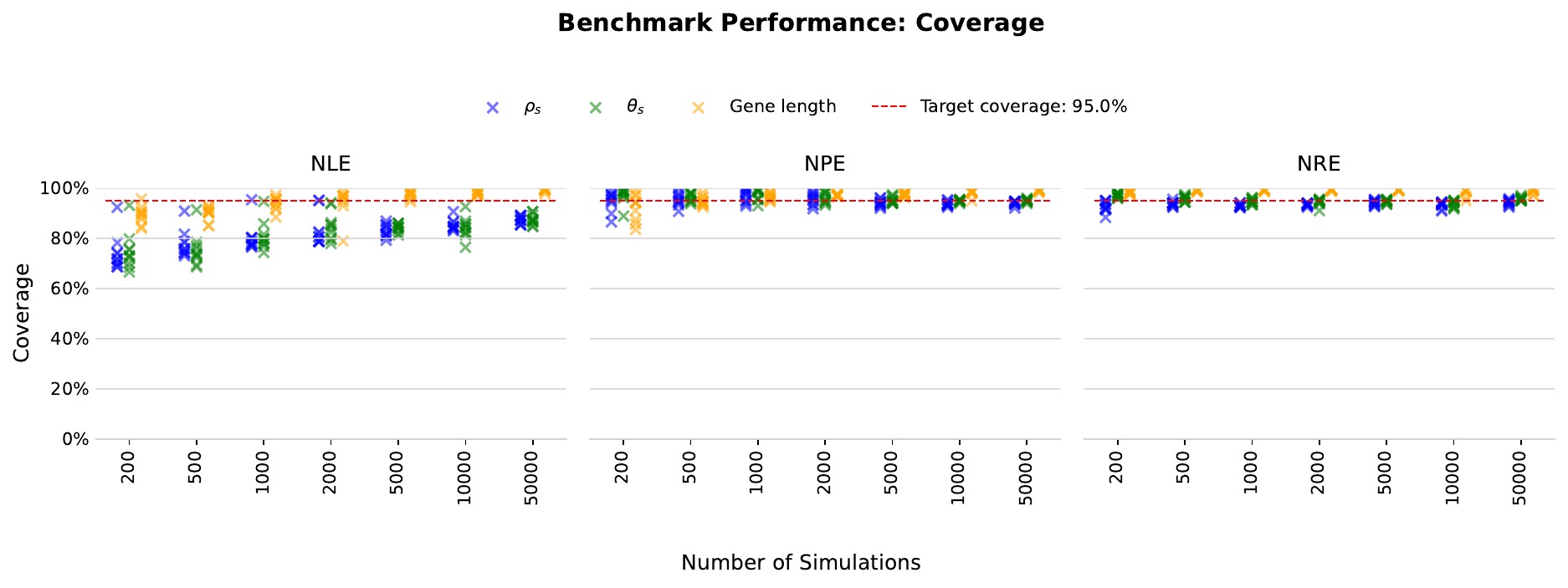}
\end{center}
\caption{Comparison of NPE, NLE, and NRE across training simulation budgets based on the empirical coverage of the $95\%$ posterior credible intervals. We expect the empirical coverage to stay around $95\%$ for an accurate estimated posterior. This value too high might suggest underconfidence, and too low might suggest overconfidence.
\label{fig:benchmark_coverage}}
\end{figure}

\begin{figure}[H]
\begin{center}
\includegraphics[width=10cm]{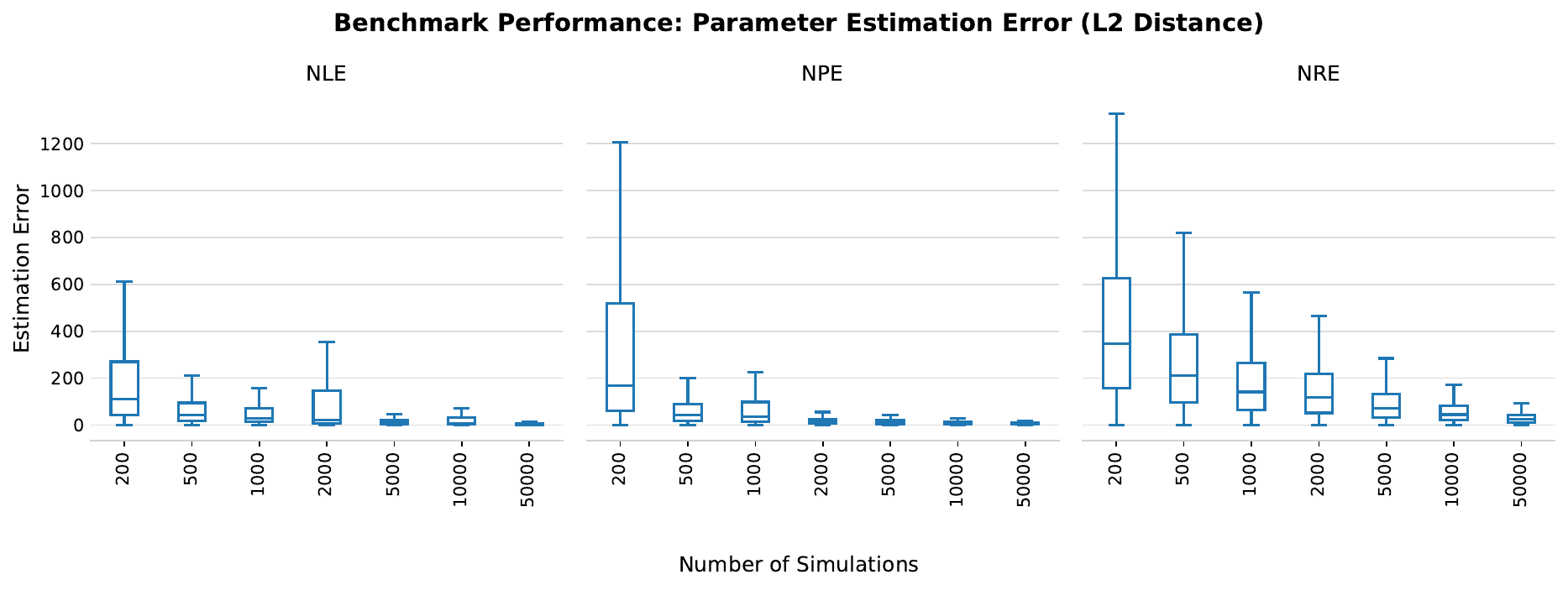}
\end{center}
\caption{Comparison of NPE, NLE, and NRE across training simulation budgets using the Euclidean ($L_2$) distance between the posterior medians and the corresponding true parameter values. Lower values indicate greater estimation accuracy.
\label{fig:benchmark_error}}
\end{figure}

To perform the benchmarking comparison between NLE, NPE, and NRE, we use the 
implementations of these three methods provided by the \texttt{sbi} package \citep{BoeltsDeistler_sbi_2025} with default model configurations. We use the simulation data for training and evaluation from Section~\ref{sec:application_clonalorigin_sim}.
To ensure a consistent comparison, all three methods were trained using their default model configurations without learned embedding networks. The training simulation budget was progressively increased from $200$ to $50000$, while the same set of $1000$ independently simulated test datasets was used for evaluation at each budget.
We found NPE generally requires less simulation budgets to achieve good testing results than NLE and NRE (Supplementary Figures \ref{fig:benchmark_sbc_p}, 
\ref{fig:benchmark_tarp_p}, \ref{fig:benchmark_coverage} and \ref{fig:benchmark_error}). 
NPE is also more computationally efficient when posterior distributions were required for many observations. Hence we use NPE as the main method of inference in this paper.

\section{Model tuning for NPE}
\label{appendix:tuning}

This section describes the procedure used to select the architecture and training configuration for NPE. Hyperparameter tuning was conducted under a fixed experimental setting so that differences in performance could be attributed to model configuration rather than to changes in the simulation budget or data-generating process. All candidate models were trained on $10000$ simulations and evaluated on $1000$ simulations from Section~\ref{sec:application_clonalorigin_sim}.

\subsection{Evaluation criteria}

The primary criterion for model selection was the validation negative log-likelihood (NLL). For a validation set $\{(\params_i,z_i)\}_{i=1}^{N_{\mathrm{val}}}$, the NLL was computed as
\begin{equation}
    \mathcal{L}_{\mathrm{val}}
    = -\frac{1}{N_{\mathrm{val}}}
    \sum_{i=1}^{N_{\mathrm{val}}}
    \log q_{\phi}(\params_i \mid z_i),
\end{equation}
where $q_{\phi}(\params \mid z)=q_{\phi}(\params\mid e_{\psi_{\mathrm{e}}}(z))$ denotes the posterior density estimated by NPE. A lower validation NLL indicates that the learned posterior assigns greater probability density to the parameters that generated the validation simulations.

\subsection{Hyperparameter search space}

The search considered three components of the NPE pipeline: the input representation and compression network, the conditional normalising flow, and the optimisation settings. The candidate values are summarised in Table~\ref{tab:npe_tuning_space}.

\begin{table}[htbp]
    \centering
    \caption{Hyperparameter search space used for NPE model tuning.}
    \label{tab:npe_tuning_space}
    \begin{tabular}{p{0.29\textwidth}p{0.61\textwidth}}
        \hline
        \textbf{Component} & \textbf{Candidate values} \\
        \hline
        Input representation & Full input; exclusion of exact LD and G4 features; exclusion of proportional LD and G4 features \\
        Compression network depth & $1$, $2$, $3$, or $4$ hidden layers \\
        Hidden units per layer & $32$, $48$, $64$, $128$, or $256$ \\
        Network output dimension & $2+1$, $2\times2+1$, $2\times4+1$, or $2\times8+1$ \\
        Density estimator & Neural spline flow (NSF) or masked autoregressive flow (MAF) \\
        Number of flow transforms & $3$, $5$, or $8$ \\
        Flow hidden features & $30$, $50$, $80$, or $120$ \\
        Number of spline bins (NSF) & $5$, $10$, or $20$ \\
        Learning rate & $10^{-4}$, $5\times10^{-4}$, or $10^{-3}$ \\
        Early-stopping patience & $10$, $20$, or $25$ epochs \\
        Training batch size & $100$, $200$, or $300$ \\
        Maximum gradient norm & No clipping, $1$, $5$, or $8$ \\
        \hline
    \end{tabular}
\end{table}

The output dimensions of the compression network were parameterised relative to the two rate parameters, $\rho_s$ and $\theta_s$, with one additional dimension for the appended gene length (Section~\ref{sec:compression_network}). Thus, the candidates $2+1$, $2\times2+1$, $2\times4+1$, and $2\times8+1$ correspond to $m=2$, $4$, $8$ and $16$ learned features and to total output dimensions of $3$, $5$, $9$, and $17$, respectively. This range allowed us to assess whether a compact representation was sufficient or whether a higher-dimensional embedding improved posterior estimation.

\subsection{Sequential tuning procedure}

An exhaustive search over all combinations would have required a prohibitively large number of training runs. We therefore used a sequential procedure in which one group of hyperparameters was varied at each stage while the remaining settings were held fixed at their current baseline or previously selected values.

As an initial reference, a model without a learned compression network was trained using an NSF with five transforms, $50$ hidden features, and $10$ spline bins. The learning rate was $5\times10^{-4}$, the batch size was $200$, the early-stopping patience was $20$ epochs, and the maximum gradient norm was $5$. A baseline model was then constructed by adding a compression network with two hidden layers, $48$ units per layer, and an output dimension of $2\times2+1$; all flow and optimisation settings were otherwise unchanged.

Starting from this baseline, tuning proceeded through the following stages:
\begin{enumerate}
    \item \textbf{Input representation.} The three input variants were compared. Their performance was similar, and the full input representation was retained because it preserved the greatest amount of potentially informative information.

    \item \textbf{Compression output dimension.} The four candidate embedding dimensions were evaluated. The two smallest representations, $2+1$ and $2\times2+1$, were retained for the subsequent architecture search.

    \item \textbf{Compression network architecture.} Combinations of network depth and hidden width were screened. Each configuration was evaluated once in the broad search. The eight strongest configurations were then assessed over five repeated runs to reduce sensitivity to stochastic optimisation. Three architectures were retained: one hidden layer with $128$ units, four hidden layers with $256$ units, and four hidden layers with $48$ units. All three used the $2+1$ output representation.

    \item \textbf{Density-estimator family.} NSF and MAF density estimators were compared using the retained compression network configurations. NSF provided the stronger overall performance and was used in the remaining stages.

    \item \textbf{Normalising-flow architecture.} The number of transforms, hidden features, and spline bins were tuned jointly. A broad single-run search was followed by five repeated runs for the eight strongest configurations. The three retained combinations were: (i) one compression network layer with $128$ units, eight transforms, $30$ hidden features, and five bins; (ii) four compression network layers with $256$ units, eight transforms, $30$ hidden features, and $20$ bins; and (iii) four compression network layers with $48$ units, eight transforms, $50$ hidden features, and $10$ bins.

    \item \textbf{Optimisation settings.} Finally, the learning rate, early-stopping patience, batch size, and gradient-clipping threshold were tuned. As in the preceding architecture searches, an initial single-run screening was followed by repeated evaluation of the strongest candidates.
\end{enumerate}

This staged strategy reduced the computational cost of tuning while allowing the major architectural and optimisation choices to be evaluated systematically. Repeated training of the leading candidates was used to reduce the influence of random initialisation and minibatch order.

\subsection{Selected configuration}

The final NPE configuration is reported in Table~\ref{tab:final_npe_configuration}. The selected compression network was relatively compact, consisting of one hidden layer with $128$ units and an output dimension of $2+1=3$. The posterior density estimator was an NSF with eight transforms, $30$ hidden features, and five spline bins. Training used a learning rate of $10^{-4}$, a batch size of $100$, and an early-stopping patience of $20$ epochs. Gradient clipping was not applied, and weight decay was set to zero.

\begin{table}[htbp]
    \centering
    \caption{Final NPE architecture and training configuration.}
    \label{tab:final_npe_configuration}
    \begin{tabular}{ll}
        \hline
        \textbf{Hyperparameter} & \textbf{Selected value} \\
        \hline
        Input representation & Full input \\
        Compression network hidden layers & $1$ \\
        Hidden units & $128$ \\
        Network output dimension & $2+1=3$ \\
        Density estimator & NSF \\
        Number of transforms & $8$ \\
        Flow hidden features & $30$ \\
        Number of spline bins & $5$ \\
        Learning rate & $10^{-4}$ \\
        Early-stopping patience & $20$ epochs \\
        Training batch size & $100$ \\
        Maximum gradient norm & No clipping \\
        \hline
    \end{tabular}
\end{table}

\section{Additional results on the applications to real datasets}

\begin{figure}[H]
    \centering
    \includegraphics[width=5cm]{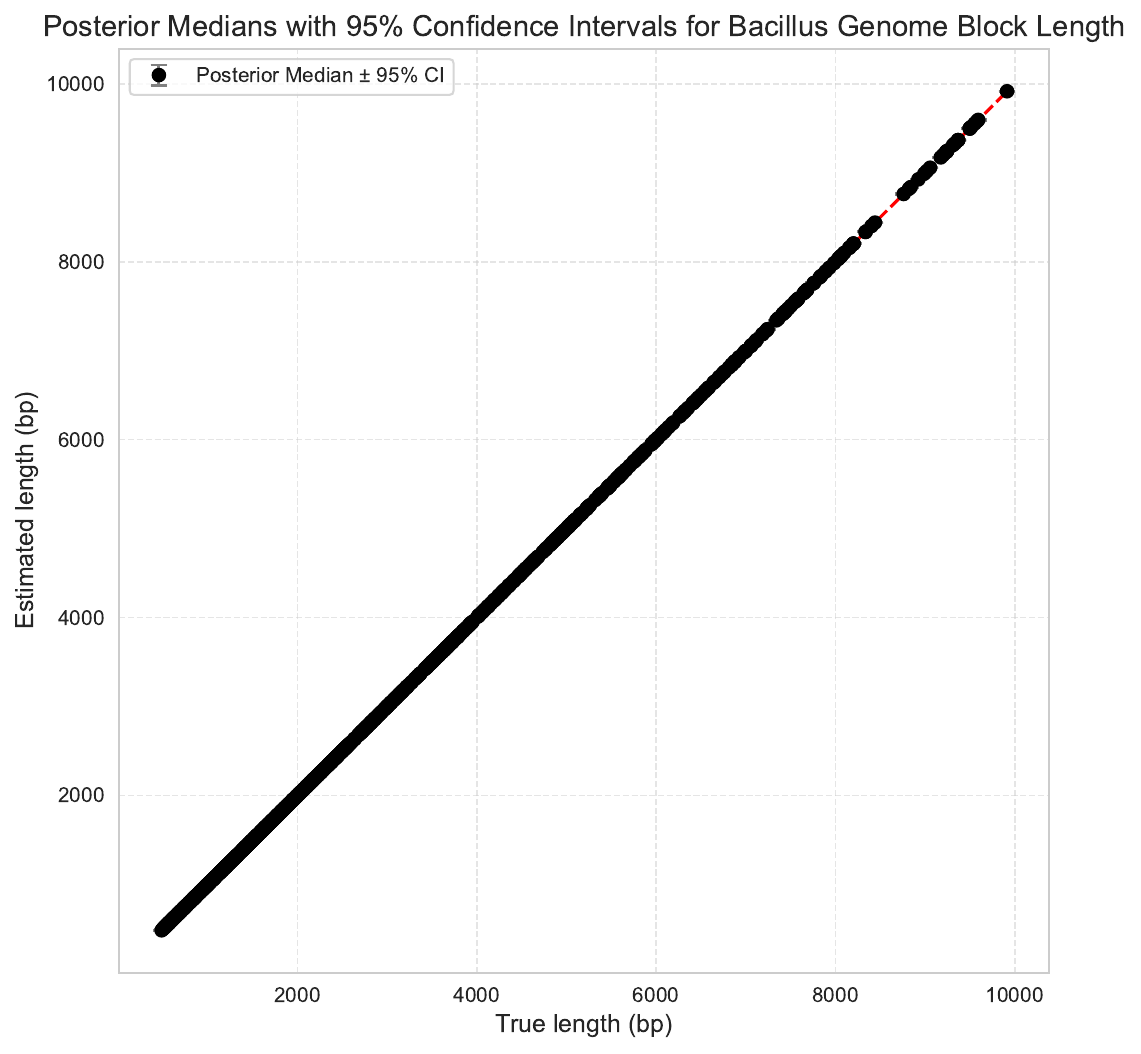}
    \caption{Comparison of the inferred and observed segment lengths for \textit{B. cereus}. Agreement between the two provides an additional diagnostic of the information retained by the trained NPE model.}
    \label{fig:bacillus_length}
\end{figure}

\begin{figure}[H]
    \centering
    \includegraphics[width=5cm]{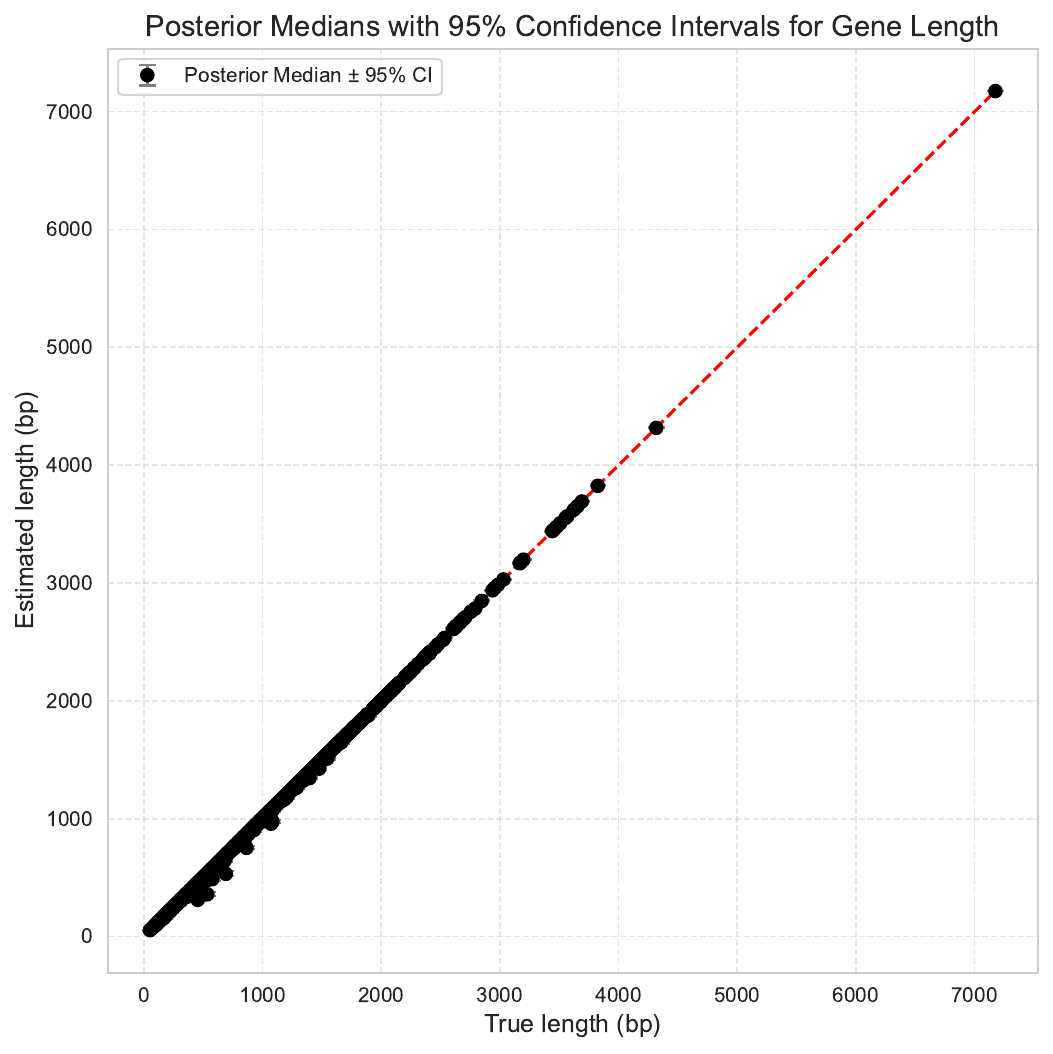}
    \caption{Comparison of the inferred and observed gene lengths for \textit{S. aureus}. Agreement between the two provides an additional diagnostic of the information retained by the trained NPE model.}
    \label{fig:staph_length}
\end{figure}

\begin{figure}[H]
    \centering
    \includegraphics[width=5cm]{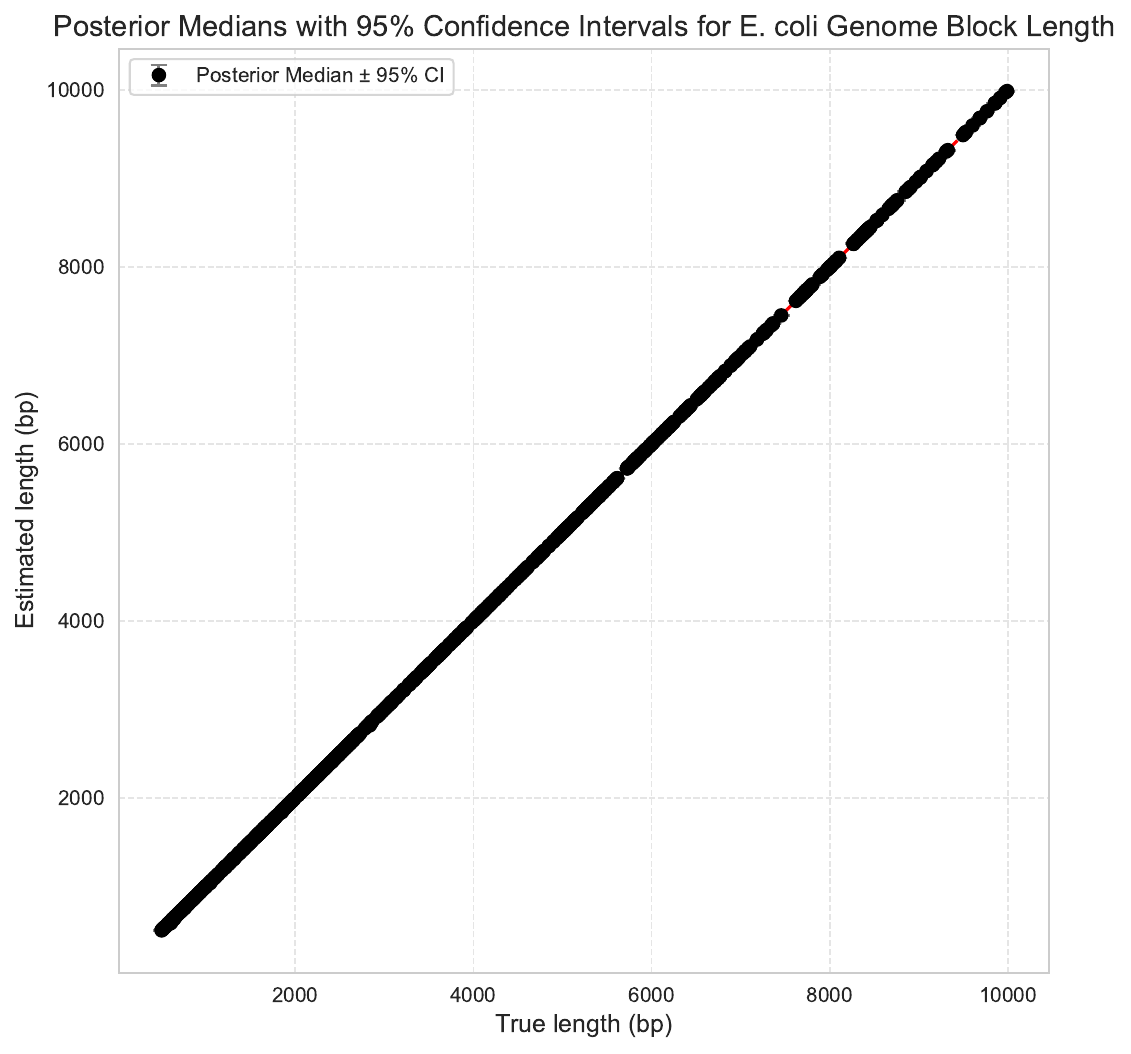}
    \caption{Comparison of the inferred and observed gene lengths for \textit{E. coli}. Agreement between the two provides an additional diagnostic of the information retained by the trained NPE model.}
    \label{fig:ecoli_length}
\end{figure}

This section reports the gene-length diagnostics (Figures~\ref{fig:bacillus_length}--\ref{fig:ecoli_length}) and the candidate hotspot regions (Tables~\ref{tab:bacillus_hotspots}--\ref{tab:ecoli_hotspots}) for the three real-data applications presented in the main text. In the tables, peak probability and peak fold enrichment are the largest values of the hotspot probability and of the fold enrichment among the genes in each region.

\begin{table}[H]
  \centering
  \begin{tabular}{cccccc}
    \hline
    \shortstack{Start position\\(bp)} &
    \shortstack{End position\\(bp)} &
    \shortstack{Number of\\genes} &
    \shortstack{Peak\\probability} &
    \shortstack{Peak fold\\enrichment} &
    Hotspot \\
    \hline
    596779  & 633202  & 16 & 1.000 & 2.31 & Recombination \\
    642338  & 672021  & 9  & 1.000 & 3.04 & Recombination \\
    819132  & 831362  & 6  & 0.968 & 1.72 & Recombination \\
    866656  & 877393  & 3  & 0.924 & 1.70 & Recombination \\
    917005  & 1033471 & 25 & 1.000 & 5.49 & Recombination \\
    1065265 & 1071176 & 3  & 0.907 & 1.72 & Recombination \\
    1790666 & 1809442 & 9  & 0.976 & 1.78 & Recombination \\
    2800266 & 2853730 & 19 & 1.000 & 2.46 & Recombination \\
    2908540 & 3012194 & 19 & 1.000 & 4.64 & Recombination \\
    3107059 & 3124701 & 8  & 1.000 & 2.97 & Recombination \\
    3147941 & 3236668 & 19 & 0.998 & 2.32 & Recombination \\
    4332066 & 4357507 & 7  & 1.000 & 1.99 & Recombination \\
    4461006 & 4468587 & 2  & 0.919 & 1.69 & Recombination \\
    4522868 & 4561439 & 12 & 1.000 & 2.34 & Recombination \\
    4566313 & 4636005 & 21 & 1.000 & 2.60 & Recombination \\
    4703349 & 4762964 & 18 & 0.999 & 1.95 & Recombination \\
    4795170 & 4841840 & 18 & 1.000 & 2.52 & Recombination \\
    \hline
  \end{tabular}
  \caption{Candidate hotspot regions detected in the \textit{Bacillus cereus} application. Positions refer to the \textit{B. anthracis} Ames 0581 reference genome.}
  \label{tab:bacillus_hotspots}
\end{table}

\begin{table}[H]
  \centering
  \begin{tabular}{cccccc}
    \hline
    \shortstack{Start position\\(bp)} &
    \shortstack{End position\\(bp)} &
    \shortstack{Number of\\genes} &
    \shortstack{Peak\\probability} &
    \shortstack{Peak fold\\enrichment} &
    Hotspot \\
    \hline
    2900271 & 20720   & 18 & 0.965 & 1.71  & Recombination \\
    576837  & 598142  & 18 & 0.969 & 2.35  & Recombination \\
    2173762 & 2189640 & 9  & 0.961 & 2.68  & Recombination \\
    2578730 & 2614375 & 26 & 1.000 & 7.13  & Recombination \\
    2637343 & 2660657 & 11 & 0.996 & 3.06  & Recombination \\
    2866243 & 2885569 & 10 & 0.942 & 1.69  & Recombination \\
    104041  & 123499  & 10 & 1.000 & 55.66 & Mutation \\
    450865  & 480302  & 8  & 1.000 & 59.07 & Mutation \\
    1931088 & 1965757 & 30 & 1.000 & 68.01 & Mutation \\
    2024563 & 2029677 & 6  & 0.930 & 46.47 & Mutation \\
    2173762 & 2190739 & 10 & 0.997 & 60.51 & Mutation \\
    2492529 & 2500566 & 10 & 0.933 & 46.62 & Mutation \\
    2522500 & 2534933 & 9  & 0.972 & 49.91 & Mutation \\
    2622988 & 2667230 & 25 & 0.999 & 55.90 & Mutation \\
    2781389 & 2790572 & 6  & 0.931 & 45.58 & Mutation \\
    2843633 & 2849825 & 8  & 0.912 & 44.24 & Mutation \\
    \hline
  \end{tabular}
    \caption{Candidate hotspot regions detected in the \textit{Staphylococcus aureus} application. The first region spans the origin of the circular chromosome.}
  \label{tab:staph_hotspots}
\end{table}

\begin{table}[H]
  \centering
  \begin{tabular}{cccccc}
    \hline
    \shortstack{Start position\\(bp)} &
    \shortstack{End position\\(bp)} &
    \shortstack{Number of\\genes} &
    \shortstack{Peak\\probability} &
    \shortstack{Peak fold\\enrichment} &
    Hotspot \\
    \hline
    1844174 & 2071008 & 22 & 1.000 & 7.57 & Recombination \\
    2791496 & 2916774 & 20 & 1.000 & 4.61 & Recombination \\
    1844174 & 2073314 & 24 & 1.000 & 342.92 & Mutation \\
    4676291 & 4728841 & 19 & 1.000 & 342.88 & Mutation \\
    \hline
  \end{tabular}
  \caption{Candidate hotspot regions detected in the \textit{Escherichia coli} application. Positions refer to the \textit{E. coli} strain 536 reference genome.}
  \label{tab:ecoli_hotspots}
\end{table}

\newpage
\printbibliography
\end{refsection}

\end{document}